\documentclass[prd,twocolumn,reprint,preprintnumbers,nofootinbib]{revtex4-1}

\input{header}

\begin{document}
\title{Looking forward to test the KOTO anomaly with FASER}

\author{Felix Kling}
\email{felixk@slac.stanford.edu}
\affiliation{SLAC National Accelerator Laboratory, 2575 Sand Hill Road, Menlo Park, CA 94025, USA}

\author{\vspace{-0.2cm}Sebastian~Trojanowski}
\email{s.trojanowski@sheffield.ac.uk}
\affiliation{Consortium for Fundamental Physics, School of Mathematics and  Statistics, University of Sheffield, Hounsfield Road, Sheffield, S3 7RH, UK}
\affiliation{National Centre for Nuclear Research, Pasteura 7, 02-093 Warsaw, Poland}

\begin{abstract}
The search for light and long-lived particles at the LHC will be intensified in the upcoming years with a prominent role of the new FASER experiment. In this study, we discuss how FASER could independently probe such scenarios relevant for new physics searches at kaon factories. We put an emphasis on the proposed explanations for the recently observed three anomalous events in the KOTO experiment. The baseline of FASER precisely corresponds to the proposed lifetime solution to the anomaly that avoids the NA62 bounds on charged kaons. As a result, the experiment can start constraining relevant models within the first few weeks of its operation. In some cases, it can confirm a possible discovery with up to $10^4$ spectacular high-energy events in FASER during LHC Run 3. Further complementarities between FASER and kaon factories, which employ FASER capability to study $\gamma\gamma$ signatures, are illustrated for the model with axion-like particles dominantly coupled to $SU(2)_W$ gauge bosons.
\end{abstract}

\maketitle

\section{Introduction}

The quest for beyond the Standard Model (BSM) physics is undoubtedly among the biggest challenges in contemporary physics. It has typically been driven by the lack of answers to some fundamental questions, such as understanding the unification of the forces of nature or the mechanism of baryogenesis, as well as by hints of new physics appearing in anomalous experimental observations. The latter become especially intriguing when the known Standard Model (SM) background (BG) can be reduced to negligible levels so that even the observation of a few events can be attributed to BSM effects.

The flavor-changing rare decays of kaons are exceptionally promising examples of such discovery channels given their highly suppressed SM rates. In particular, the KOTO collaboration has recently reported a possible observation of three anomalous events in the search for $K_L\to\pi^0\nu\bar{\nu}$~\cite{KOTOanomaly}, in which no SM signal was expected at the current level of the experimental sensitivity. If taken as a signal, this observation would indicate that the relevant branching fraction exceeds the SM prediction by about two orders of magnitude~\cite{Kitahara:2019lws}. It has been quickly noticed that the discrepancy could be resolved by introducing a new light long-lived particle (LLP) with mass of order $100~\mev$, light enough to be produced in kaon decays. By properly adjusting the lifetime of the LLP, one can explain the neutral kaon anomaly while avoiding stringent bounds on such models from searches for charged kaon decays. 

In this study, we illustrate how such proposed solutions to the KOTO anomaly could be thoroughly probed by the recently approved ForwArd Search ExpeRiment, or FASER~\cite{Feng:2017uoz, Ariga:2018zuc, Ariga:2018pin}. The FASER experiment will search for LLPs abundantly produced in the far-forward direction of the Large Hadron Collider (LHC) during its upcoming Run 3. Several features of its setup make it an ideal probe for light LLPs produced in kaon decays: 
i) Given a large expected boost factor of light particles traveling towards FASER, and the location of the detector, the experiment's sensitivity is optimal for the precise combination of the lifetime and mass that are required to fit the anomaly. In particular, for high-energy LLPs with $E\sim 1~\tev$ that are produced at the ATLAS IP, the sensitivity reach of FASER is maximized for
\begin{equation}
\frac{\tau_{\textrm{LLP}}}{m_{\textrm{LLP}}} \sim \frac{0.1~\textrm{ns}}{100~\mev},\hspace{1cm}\textrm{(best FASER reach)}.
\end{equation}
ii) The FASER sensitivity additionally benefits from the $\mathcal{O}(150~\m)$ long part of the LHC beam pipe, in which forward-going neutral kaons can travel and decay before being absorbed, as well as the presence of additional LLP production modes at the LHC. 
iii) The FASER detector has the capability to not only detect LLP decays into charged particles but also into photon pairs, due to its dedicated pre-shower detector. As a result, during Run 3 of the LHC, FASER can turn the three currently observed anomalous events at KOTO into up to $\mathcal{O}(10^4)$ spectacular high-energy LLP decay events with expected negligible BG.

The rest of this paper is organized as follows. In \cref{sec:KOTO} and \cref{sec:FASER} we provide more details about the KOTO anomaly and the FASER experiment, respectively. The FASER sensitivity reach in several selected BSM scenarios is presented in \cref{sec:results}. In \cref{sec:ALPs}, we further illustrate the possible interplay between FASER and searches for LLPs in kaon factories. We conclude in \cref{sec:conclusion}. Several useful expressions for BSM meson decay branching fractions are listed in \cref{app:BRs}. 

\section{\label{sec:KOTO} KOTO experiment and the anomaly}

The KOTO experiment~\cite{Yamanaka:2012yma} has been designed to study decays of kaons into neutral pions and a neutrino/anti-neutrino pair, $K_L\rightarrow\pi^0\nu\bar{\nu}$. The kaons are produced in collisions of $30~\gev$ protons from the Japan Proton Accelerator Research Complex~\cite{Nagamiya:2012tma} (J-PARC) main ring accelerator with the target made out of gold. A high-intensity beam of neutral kaons produced at an angle of $16^\circ$ with respect to the proton beam line is created with the use of dedicated collimators. At a distance of about $24~\m$ away from the target, the $\sim\!3$ m long vacuum chamber of the KOTO detector begins. Here, kaon decays are identified by detecting photons produced in prompt neutral pion decays.

The data collected by the KOTO collaboration in 2015 allowed them to obtain the leading bound on the branching fraction of the aforementioned decay process. At the $90\%$ confidence level (CL) it reads~\cite{Ahn:2018mvc}
\begin{equation}
\mathcal{B}_{\textrm{KOTO,bound}}(K_L\rightarrow\pi^0\nu\bar{\nu}) \leq 3\times 10^{-9}.
\label{eq:K0Lneutbound}
\end{equation}
Notably, \cref{eq:K0Lneutbound} is consistent with the SM prediction which remains about two orders of magnitude lower, $\mathcal{B}_\textrm{SM}(K_L\rightarrow\pi^0\nu\bar{\nu})= (3.4\pm 0.6)\times 10^{-11}$~\cite{Buras:2015qea}. A similar bound on the two-body decay $K^0\rightarrow\pi^0\,X$ has also been obtained, where $X$ is a stable or long-lived BSM bosonic particle,
\begin{equation}
\mathcal{B}_{\textrm{KOTO,bound}}(K_L\rightarrow\pi^0\,X) \lesssim 2.4\times 10^{-9}.
\end{equation}

In the subsequent analysis of the data from 2016-18~\cite{KOTOanomaly}, however, a total of four candidate events were found, only one of which had the properties of BG. If taken as a signal, the remaining three anomalous events point to the branching fraction which significantly exceeds the SM prediction. At the $95\%$ CL it is given by~\cite{Kitahara:2019lws} 
\begin{equation}
\mathcal{B}_{\textrm{KOTO,anom.}}(K_L\rightarrow\pi^0\nu\bar{\nu}) = (2.1^{+4.1}_{-1.7})\times 10^{-9}.
\label{eq:K0Lneutanomaly}
\end{equation}

Importantly, in the SM the aforementioned neutral kaon decay mode is also related by the isospin symmetry to the value of the branching fraction of a charged kaon decay into a pion $\pi^+$ and a neutrino/anti-neutrino pair, which proceeds via the same parton level process $s\to d\nu\bar{\nu}$. The relevant so-called Grossman-Nir (GN) bound~\cite{Grossman:1997sk} reads 
\begin{equation}
\mathcal{B}(K_L\rightarrow\pi^0\nu\bar{\nu}) < 4.3\,\mathcal{B}(K^+\rightarrow\pi^+\nu\bar{\nu}).
\label{eq:GNbound}
\end{equation}
The current upper limit on the $K^+$ decay branching fraction is obtained from the results of the E949 experiment, which observed several charged kaon decay events~\cite{Artamonov:2008qb,Artamonov:2009sz}, and it is given by $\mathcal{B}_{\textrm{E949,bound}}(K^+\rightarrow\pi^+\nu\bar{\nu})<3.35\times 10^{-10}$ at $90\%$ CL. This is consistent with the SM expectation $\mathcal{B}_\textrm{SM}(K^+\rightarrow\pi^+\nu\bar{\nu}) = (8.4\pm 1)\times 10^{-11}$~\cite{Buras:2015qea}. An improved analysis has also recently been preliminarily presented by the NA62 collaboration with the corresponding branching fraction measured to be at the $68\%$ CL~\cite{NA62bound},
\begin{equation}
\mathcal{B}_{\textrm{NA62}}(K^+\rightarrow\pi^+\nu\bar{\nu}) = 0.47^{+0.72}_{-0.47}\times 10^{-10},
\label{eq:NA62bound}
\end{equation}
and the upper bound at the $95\%$ CL given by $\mathcal{B}_{\textrm{NA62,bound}}(K^+\rightarrow\pi^+\nu\bar{\nu})<2.44\times 10^{-10}$. The measurements of the charged kaon decay branching fraction, together with the GN bound \cref{eq:GNbound}, lead to $(2-3)\sigma$ tension with the anomalous observation by the KOTO collaboration, cf. \cref{eq:K0Lneutanomaly}, depending on how the possible impact of new physics is taken into account~\cite{Kitahara:2019lws}. A number of BSM scenarios have been proposed to explain this discrepancy~\cite{Kitahara:2019lws, Fabbrichesi:2019bmo, Egana-Ugrinovic:2019wzj, Li:2019fhz,Dev:2019hho, Liu:2020qgx,Jho:2020jsa, Cline:2020mdt,He:2020jzn, Ziegler:2020ize, Liao:2020boe, He:2020jly, Gori:2020xvq, Hostert:2020gou, Datta:2020auq, Dutta:2020scq,Altmannshofer:2020pjb,Liu:2020ser, Archer-Smith:2020hqq}.

Among these scenarios, a prominent role is played by models predicting the existence of a new LLP with the mass $m_X\sim 100~\mev$, which can be produced in rare kaon decays, $K_L\rightarrow\pi^0\,X$~\cite{Kitahara:2019lws}. In particular, if such a light BSM particle has a lifetime of order $\tau_X\sim 0.1$~ns, it can be effectively stable and invisible in the search performed by the KOTO collaboration. At the same time, $X$ will typically decay inside the E949 and NA62 detectors, and it does not contribute to the measured branching fraction of $K^+_L\rightarrow\pi^+\nu\bar{\nu}$. This leads to an apparent violation of the GN bound when the results of these experiments are compared with each other, which is further referred to as the \textsl{lifetime solution} to the anomaly. 

A very long-lived $X$ can also explain the anomaly while avoiding the bounds from past beam-dump searches. In addition, if its mass is close to the pion mass, it can escape detection in E949 and NA62 experiments by hiding in the SM BG~\cite{Fuyuto:2014cya}. Another distinct possibility is to allow $K_L$ decays into neutral dark states, $K_L\rightarrow X_1\,X_2$. Subsequent $X_i$ decays that produce photons can be detected in the KOTO electromagnetic calorimeter and resemble a neutral pion signature~\cite{Hostert:2020gou}. In this case, charged kaon three-body decays can be suppressed or even kinematically forbidden. 

Below, we analyze the prospect of probing such selected BSM scenarios in the FASER experiment at the LHC.

\section{\label{sec:FASER}FASER and FASER~2 experiments}

The FASER experiment~\cite{Feng:2017uoz,Ariga:2018zuc,Ariga:2018pin} has been proposed to search for LLPs~\cite{Beacham:2019nyx, Alimena:2019zri} produced in the forward region of the LHC~\cite{Feng:2017vli, Kling:2018wct, Feng:2018noy, Ariga:2018uku, Berlin:2018jbm, Ariga:2019ufm, Jodlowski:2019ycu}, as well as to study interactions of high-energy neutrinos~\cite{Abreu:2019yak, Abreu:2020ddv, Kling:2020iar}. It utilizes the fact that light and high-energy particles produced in $pp$ collisions at the LHC and e.g. subsequent decays of light mesons $M$, will typically travel in the forward direction, as dictated by their estimated angular spread around the beam collision axis, $\theta_X\sim m_M/E_X\ll 1$. As a result, even a small detector placed along the beam collision axis can search for displaced decays of LLPs, provided that it is shielded to suppress the SM BG. 
The FASER detector will operate during the LHC Run 3 in the former service tunnel TI12 located at a distance $L=480~\m$ away from the ATLAS IP. Below, we also show the sensitivity reach for the same detector with continued operation during the High-Luminosity LHC phase (HL-LHC). We refer to this version of the experiment as FASER~HL. We also present the results obtained for a possible larger version of the detector to operate during HL-LHC, dubbed FASER~2. In particular, we assume a cylindrical decay volume with length $\Delta$, radius $R$, and an integrated luminosity $\mathcal{L}$ for each of the three aforementioned cases:
\be
\textrm{\textbf{FASER:}}  
&\ \Delta = 1.5~\m ,
\ R=10~\cm,
\ \mathcal{L} = 150~\ifb,\nonumber\\
\textrm{\textbf{FASER~HL:}}  
&\ \Delta = 1.5~\m,
\ R=10~\cm,
\ \mathcal{L} = 3~\iab,\nonumber\\
\textrm{\textbf{FASER~2:}} 
&\ \Delta = 5~\m,
\ R=1~\m, \phantom{0c}
\ \mathcal{L} = 3~\iab.\nonumber
\ee
Our main focus in this study is to highlight possible complementarity between FASER searches for LLPs and kaon factories, with a particular emphasis on the recently observed KOTO anomaly. It is then useful to briefly summarize the main advantages of FASER that make it an ideal tool to study related BSM scenarios predicting a new unstable light species:
\begin{itemize}
\item FASER has the capability to study di-photon final states in LLP decays, on top of more often considered electron-positron pairs.
\item Forward-going kaons produced in $pp$ collisions at the ATLAS IP can travel about $140~\m$, and hence decay with sizable probability, before they are absorbed. Additional production modes can further improve detection prospects.
\item FASER's sensitivity is optimal for typical LLP mass and lifetime proposed to explain the KOTO anomaly, cf. discussion in \cref{sec:KOTO}.
\end{itemize}

The signatures that we focus on consist of LLP $X$ decays into mainly $\gamma\gamma$ or $e^+e^-$ pairs that can be resolved in the detector. For sufficiently large energy, which we take to be $E_{X}\gtrsim 100~\gev$, the search at the FASER location can be considered BG-free~\cite{Ariga:2018zuc,Ariga:2018pin}. In the following, we assume a $100\%$ detection efficiency. Notably, the preliminary efficiency studies show that it typically has a minor impact on the sensitivity reach plots~\cite{Ariga:2018zuc}. 

\paragraph{FASER capabilities to study the $\gamma\gamma$ final state} As far as the $\gamma\gamma$ final state is concerned, a very good separation efficiency between the photons is achieved thanks to a dedicated pre-shower detector placed in front of the FASER calorimeter, i.e. $2~\m$ after the end of the FASER decay vessel. A preliminary relevant discussion can be found in Ref.~\cite{Feng:2018noy} in the context of FASER search for axion-like particles with the dominant di-photon coupling. As shown there, taking into account a finite separation resolution between the two $\gamma$s of order $\delta\sim 200$~nm, which is achievable in FASER, has a negligible impact on the sensitivity reach. Instead, the substantial effect was expected only if the resolution was worse than about $1$~mm. 

Additionally, LLP decays into photons could also be proved in the single photon channel. Importantly, single high-energy BG photons with $E_\gamma\gtrsim 100~\gev$ at FASER location are typically associated with time-coincident muons activating veto layers. Hence, a large number of very energetic photon pairs produced in LLP decays even closer to the pre-shower detector, which could be misidentified as a single EM shower, would still be indicative of new physics.

\paragraph{Forward going kaons at the LHC} It is also important to stress that forward-going kaons produced at the ATLAS IP will not be immediately absorbed. Instead, they can travel a long distance through the LHC beam pipe, which allows them to decay with a sizable probability. In particular, for neutral kaons produced within the angular coverage of FASER, $\theta_K < \theta_{\textrm{FASER}}\approx 0.2~\mrad$, the closest element of the LHC infrastructure that they hit is the TAN neutral particle absorber placed at a distance $140~\m$ away from the ATLAS IP. We take this into account in our modeling along with the TAS absorber placed $20~\m$ away from the IP that affects neutral kaons produced with larger $\theta_K$. The TAS also marks the end of the region where charged kaons are not deflected away by strong LHC magnets. 

Below, we also illustrate how in specific models additional production modes of LLPs, for example in heavy meson decay, can further improve FASER sensitivity.

\paragraph{Lifetime regime} Last but not least, it is worth highlighting that in order for the LLP with $m_X\sim 100~\mev$ and energy $E_X\sim \tev$ to reach FASER, the required lifetime is of order $\tau_X\sim L/(\gamma_X\,c) \sim 0.1$~ns. This precisely corresponds to the sweet spot between the KOTO and NA62 searches discussed in \cref{sec:KOTO}. Therefore, as we will see below, it is not a surprise that FASER can effectively exclude many such explanations of the anomaly even with the first $10~\ifb$ of data. Instead, in the case of discovery, FASER can confirm the anomaly with large statistics of related LLP decay events. This can reach up to $\mathcal{O}(10^4)$ events in FASER during LHC Run 3, while it could grow even larger for FASER~HL and FASER~2.

\section{\label{sec:results}FASER sensitivity reach}

As discussed in \cref{sec:KOTO}, if the anomalous neutral kaon decay events observed in the KOTO experiment are confirmed, this would require a BSM explanation. In the following, we illustrate how such models can be independently probed in FASER. To this end, we focus on several scenarios predicting LLPs.  

\subsection{\label{sec:genericX} Generic LLP and the lifetime solution}

\begin{figure*}[tb]
\centering
\includegraphics[width=0.49\textwidth]{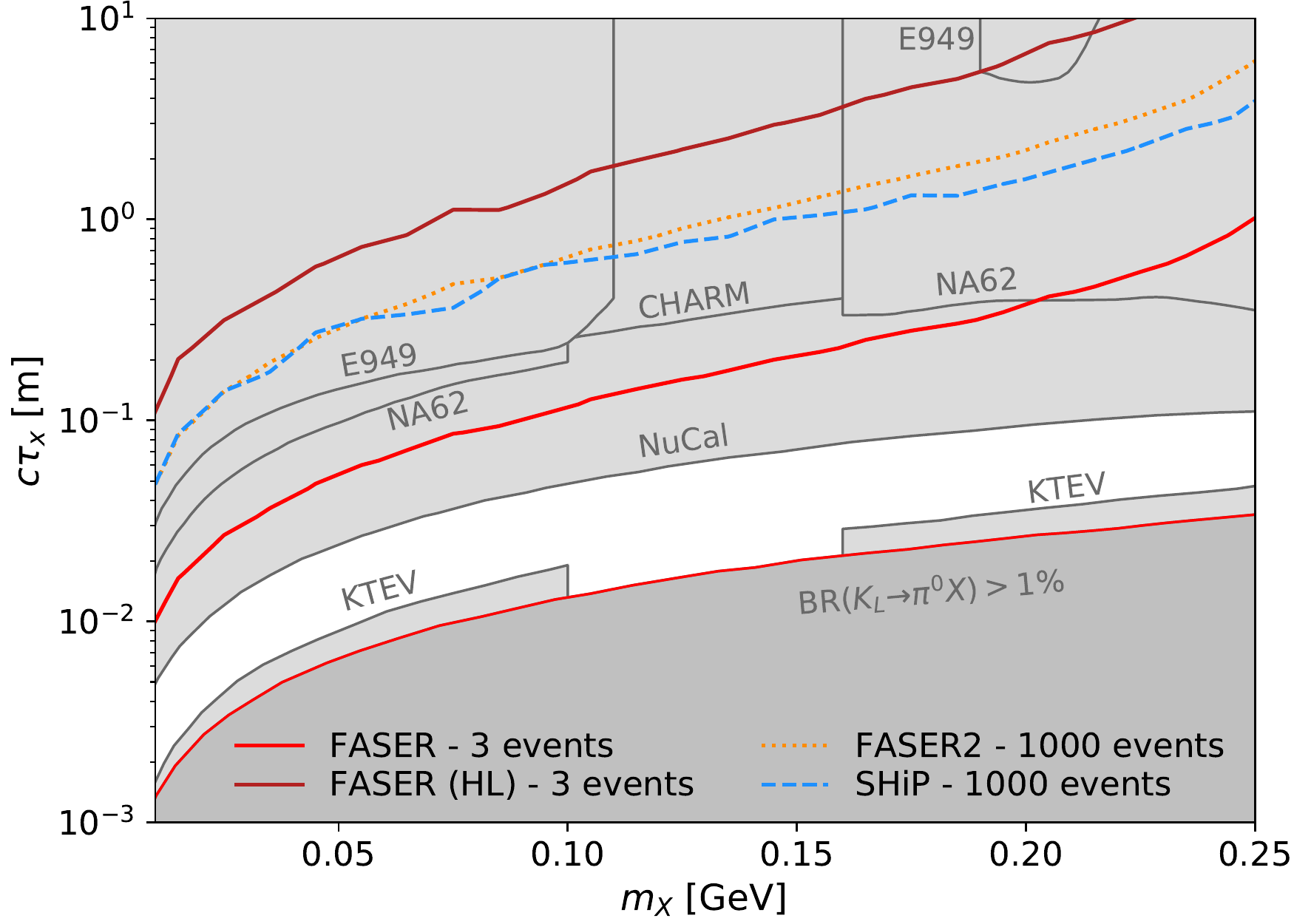}
\hfill
\includegraphics[width=0.49\textwidth]{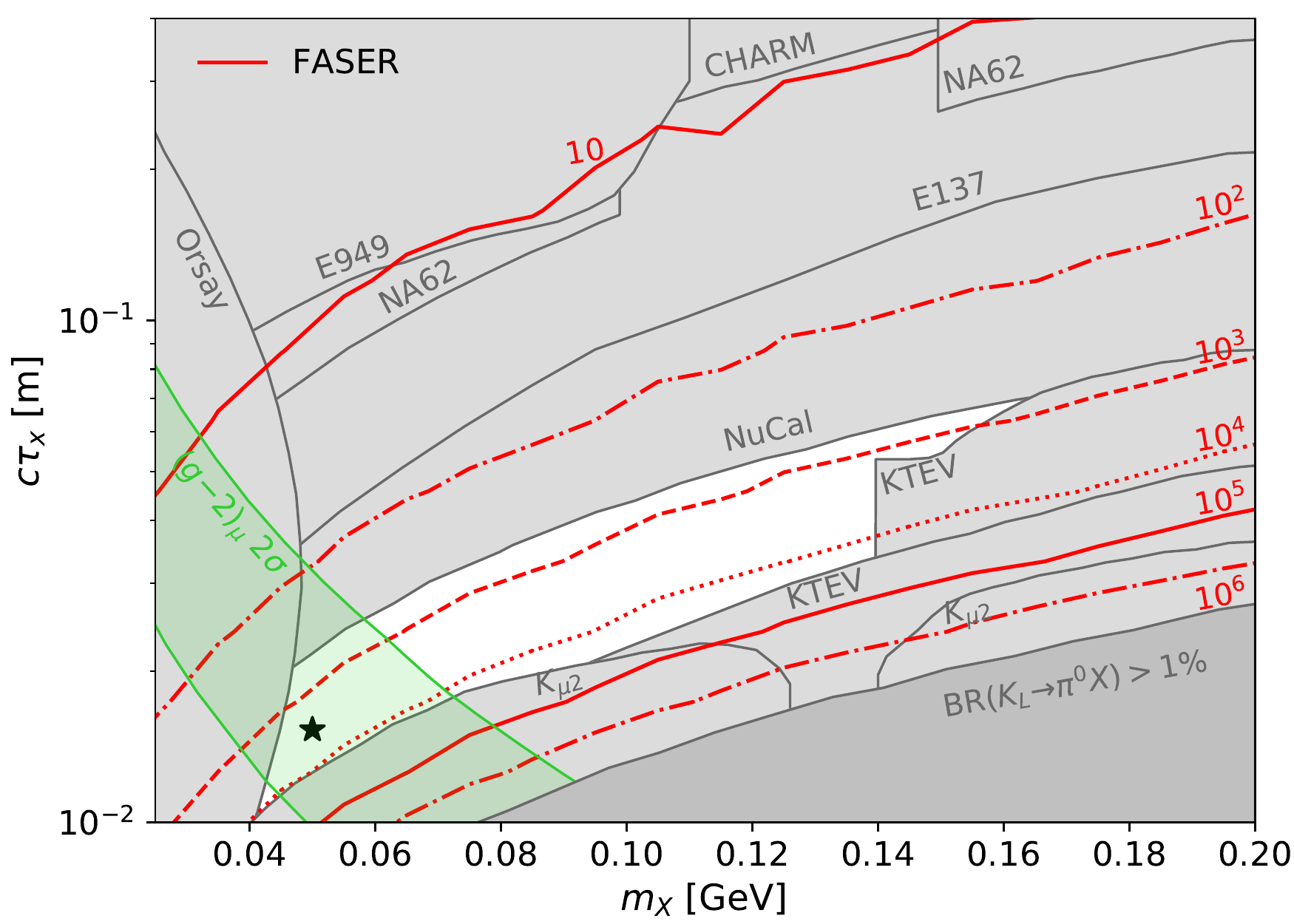}
\caption{Results for the \textsl{lifetime solution} to the KOTO anomaly (white regions in the plots) presented in the $(m_X,c\tau_X)$ plane, where $m_X$ and $\tau_X$ correspond to the LLP mass and lifetime, respectively. The current bounds are shown as gray-shaded regions. 
\textbf{Left:} Reach for the generic X scenario in which a new light and unstable particle is produced specifically in the $K_L$ and $K^+$ decays, cf. discussion in \cref{sec:genericX}. From top to bottom, dark and light red solid lines correspond to the upper limit in $c\tau_X$ of the sensitivity reach of the FASER~HL and FASER experiments, respectively. The experiments can probe scenarios relevant for the entire white region in the plot below the sensitivity lines. Also shown, are the lines with a fixed number of events $N_{\textrm{ev}}=1000$ expected for the FASER~2 (orange dotted line) and SHiP (blue dashed line) experiments. The number of events in both cases grow for lower values of $c\tau_X$ in the white region of the plot. At the bottom of the plot, the dark red solid line marks the region of the parameter space which is excluded by the untagged $K_L \to \pi^0 X $ branching fraction. 
\textbf{Right:} Event contours (from top to bottom: $10$ (solid), $10^2$ (dash-dotted), $10^3$ (dashed), $10^4$ (dotted), $10^5$ (solid) and $10^6$ (dash-dotted)) at FASER at the LHC Run 3 for the dark Higgs boson model with non-universal couplings to the SM leptons and quarks, as discussed in \cref{sec:2HDM}. The green-shaded band corresponds to the $2\sigma$ fit to the $(g-2)_\mu$ anomaly.
}
\label{fig:genericandspecificX}
\end{figure*}

The simplest BSM scenario proposed to explain the anomaly and to avoid stringent bounds on rare decays of charged kaons, employs a new unstable $X$ particle produced in kaon decays, $K_L\rightarrow \pi^0\,X$, and $K^+\rightarrow \pi^+\,X$. In particular, if the lifetime of $X$ and its mass are of order $\tau_X\sim 0.1$~ns and $m_X\sim 100~\mev$, respectively, then $X$ typically decays within the volume of the E949 and NA62 detectors and does not contribute to the measured value of $\mathcal{B}(K^+\rightarrow\pi^+\nu\bar{\nu})$. On the other hand, the probability for $X$ to escape the KOTO decay volume without decaying remains larger~\cite{Kitahara:2019lws}. 

In the left panel of \cref{fig:genericandspecificX}, we show the FASER and FASER~HL sensitivity reach contours in a generic scenario, in which $X$ is produced only in kaon  decays. We assume that the real branching fraction of a two-body neutral kaon decay is of order $\mathcal{B}(K^0\rightarrow\pi^0 X)\sim 10^{-8}-10^{-6}$. Its precise value is chosen depending on $m_X$ and $\tau_X$, as well as the KOTO detector efficiency~\cite{Ahn:2018mvc}, as detailed in Ref.~\cite{Kitahara:2019lws,Liu:2020qgx}. This is done by requiring that the measured branching fraction fits the KOTO anomaly, cf. \cref{eq:K0Lneutanomaly}, after correcting for finite escape probability from the detector. In the plot, it is also assumed that the charged kaon decay branching fraction saturates the GN bound, cf. \cref{eq:GNbound}. The gray-shaded region is excluded by the constraints from the aforementioned searches for $K^+\rightarrow\pi^+\nu\bar{\nu}$ at E949~\cite{Artamonov:2009sz} and NA62~\cite{NA62bound}, the untagged $K_L \to \pi^0 X $ branching fraction~\cite{Tanabashi:2018oca}, beam-dump searches at CHARM~\cite{Bergsma:1985qz} and NuCal~\cite{Blumlein:1990ay,Blumlein:1991xh} experiments that we implement following Refs.~\cite{Liu:2020qgx,Egana-Ugrinovic:2019wzj}, and the measurement of the branching fraction for $K_L \rightarrow \pi^0 \gamma\gamma $ by KTEV~\cite{Abouzaid:2008xm}, assuming that $X$ decays dominantly into a $\gamma\gamma$ final state.\footnote{In particular, the NuCal bounds shown in \cref{fig:genericandspecificX} have been obtained following a ``conservative'' approach discussed in Ref.~\cite{Egana-Ugrinovic:2019wzj} and by taking into account $X$ production in decays of kaons (left panel) or all the relevant mesons (right panel). We note that the bounds from hadronic beam dumps are subject to uncertainties in modeling of the high-energy tail of the meson spectrum in the target.}

The anomalous KOTO events can be fitted within the remaining portion of the parameter space shown in \cref{fig:genericandspecificX}. Notably, the reconstructed transverse momentum of the pion can more easily resemble the observed such distribution by KOTO when $m_X\lesssim 180~\mev$~\cite{Kitahara:2019lws}, or even $m_X\lesssim m_\pi$~\cite{Liao:2020boe}.

In order to obtain the FASER sensitivity reach, we generate the meson spectra using \texttt{EPOS-LHC}~\cite{Pierog:2013ria} as implemented in the simulation package \texttt{CRMC}~\cite{CRMC} and subsequently decay the mesons using the branching fractions obtained as discussed above. As can be seen, already the FASER experiment operating during the LHC Run 3 can already cover the entire region of the parameter space corresponding to the anomaly. The relevant expected number of events can be as large as $\mathcal{O}(10^2)$ for FASER, and grow by an additional factor of $20$ and a few hundred for FASER~HL and FASER~2, respectively. Notably, thanks to the use of the dedicated pre-shower detector, FASER can probe this scenario even for leptophobic $X$ which decays dominantly into a di-photon pair, as assumed in the bounds shown in the left panel of \cref{fig:genericandspecificX}.

The lifetime solution to the KOTO anomaly can also be explored by other experiments dedicated to LLP searches, although this can be limited by the lack of di-photon signal detection capabilities. We illustrate this in the left panel of \cref{fig:genericandspecificX} for the proposed SHiP experiment~\cite{Alekhin:2015byh} to operate in a similar time frame to FASER~HL and FASER~2. The sensitivity of SHiP is analyzed by conservatively requiring that the kaons decay within one nuclear interaction length in the SHiP target made out of molybdenum, $\lambda = 15.25~\cm$. We show the relevant contour line corresponding to a fixed number of expected events, $N_{\textrm{ev}}=1000$, similarly to the FASER~2 line also shown in the plot. The number of events grows larger for lower values of the LLP lifetime within the allowed region in the $(m_X,c\tau_X)$ plane. Interestingly, although kaon absorption in the target at distances greater than $\lambda$ could limit SHiP sensitivity, a large number of protons on target (POT), $N_{\textrm{POT}} = 2\times 10^{20}$, as well as the size of the decay volume, $\Delta \simeq 50~\m$, sufficiently compensate for this effect. The number of expected events in SHiP could be further enhanced once more detailed modeling of the kaon propagation in the target is performed.

\subsection{\label{sec:2HDM}Specific example employing 2HDM}

\begin{table*}
\centering
\renewcommand{\arraystretch}{1.3}
\begin{tabular}{|c||c|c||c|c|c|c|c|}
\hline
\hline
Experiment & \multicolumn{2}{c||}{Benchmark} & \multicolumn{4}{c|}{$X$ prod. in meson decays and decay in FASER} & Proton brem \\
\cline{2-8}
\cline{2-8}
&$m_X$ [MeV] & $c\tau_X [\textrm{cm}]$ & $K_L\to\pi^0 X$ & $K^+\to\pi^+ X$ & $\eta,\eta^\prime,K_S\to\pi^0 X$ & $b\to s X$ & $pp\to X+\ldots$\\
\hline
\hline
 FASER &
 \multirow{2}{*}{50} & \multirow{2}{*}{1.5} & 
 150& 50& 650& 2000&
 135\\
\cline{4-8}
 FASER~2 &
  &  & 
 80k & 30k & 250k & 11M &
 16k\\
\hline
\hline
\end{tabular}
\caption{The number of $X$ decay events in FASER (FASER~2) and the integrated luminosity of $\mathcal{L} = 150~\ifb$ ($\mathcal{L} = 3~\iab$) for various production modes of the LLP. The number of events for FASER~HL are $20$ times larger than for FASER during LHC Run 3. The results correspond to the model discussed in \cref{sec:2HDM} and to the benchmark values of the model parameters given in the text and indicated in the table. The scenario can explain both the KOTO and $(g-2)_\mu$ anomalies.
\label{tab:productionmodes}}
\end{table*}

Once a specific model of the LLP that corresponds to the aforementioned lifetime solution is considered, typically more production and decay modes appear for $X$ that should be taken into account. This has an impact on present constraints on such $X$, but can also significantly increase the expected number of events in the future detectors. We illustrate this below for FASER employing the model with a leptophilic dark scalar discussed in Ref.~\cite{Liu:2020qgx}. Interestingly, such a scalar with $40~\mev\lesssim m_X\lesssim 70~\mev$ and $\tau_X\sim 0.1$~ns could simultaneously explain the recently observed KOTO events and the measurement of the anomalous magnetic moment of the muon, $(g-2)_\mu$~\cite{Bennett:2006fi}.

The model of our interest can effectively be described by three parameters corresponding to the LLP mass, as well as its coupling constants to leptons, $\epsilon_\ell$, and quarks, $\epsilon_q$. The relevant Lagrangian reads
\be
\mathcal{L}  \supset & - m_X^2 X^2
+ \sum \epsilon_q \frac{m_q}{v} \, X \bar{q} q 
+ \sum \epsilon_\ell \frac{m_\ell}{v} \, X \bar{\ell} \ell \\
&  + \epsilon_W \frac{m_W^2}{v}\, X W^+_\mu W^{\mu -},
\label{eq:L2HDM}
\ee
where the SM Higgs boson vev is equal to $v\simeq 246~\gev$ and we set $\epsilon_W=\epsilon_q$. The couplings $\epsilon_\ell$ and $\epsilon_q$ can arise as the mixing angles in the scalar sector of the type-X two Higgs doublet model (2HDM). In particular, the model allows one to disjointly treat interactions of $X$ with the SM leptons and quarks.

Assuming a hierarchy between the couplings, $\epsilon_\ell\gg\epsilon_q$, allows one to reduce the lifetime of $X$, which is governed by the leptonic coupling $\epsilon_\ell$. Thanks to this, stringent bounds from hadronic beam-dump searches can be avoided and the kaon decay branching fraction into $X$ can be kept low, as it depends on $\epsilon_q\sim 10^{-3}-10^{-2}$.  In particular, for $\epsilon_\ell\sim \mathcal{O}(1)$ the most important bounds on long-lived $X$ come from electron beam-dump experiments E137~\cite{Bjorken:1988as} and Orsay~\cite{Davier:1989wz}. However, they do not exclude scenarios with $\tau_X\sim 0.1$~ns. 

The dominant decay mode for $X$ in the considered mass regime is into the $e^+e^-$ final state, although a loop-induced decay into the $\gamma\gamma$ pair can take values of up to $\mathcal{O}(10\%)$ of the decay branching fraction of $X$. The additional bounds from the KTEV search for $K_L\rightarrow\pi^0\,e^+e^-$ decays exclude a part of the region of the parameter space of the model with $m_X\gtrsim 140~\mev$~\cite{AlaviHarati:2003mr}, although they are not relevant for lighter $X$. We employ the decay width of $X$ following Ref.~\cite{Liu:2020qgx}.

A relative smallness of $\epsilon_q$ is consistent with only a few events currently observed in KOTO. However, as far as FASER is concerned, additional production modes of $X$ can become more important than rare kaon decays, therefore increasing the expected number of signal events. In particular, when obtaining the FASER sensitivity reach, we take into account the following production channels: 
\begin{description}
\item [Meson decays] In our modeling, we include rare two-body decays of charged and neutral long-lived kaons that propagate in the LHC beam pipe, as discussed in \cref{sec:FASER}. We also analyze the impact of short-lived kaon decays, $K_S \to \pi^0 X$. Notably, the relevant BSM branching fraction of $K_S$ is suppressed with respect to $K_L$ by an additional factor of order $\mathcal{O}(10^{-4})$ due to a larger value of the total decay width of $K_S$. On the other hand, short-lived kaons can more easily decay prior to hitting any element of the LHC forward infrastructure. 

On top of kaons, a number of other mesons can be abundantly produced in the forward direction of the LHC. In particular, we study prompt rare two-body decays $\eta\to\pi^0 X$ and $\eta^\prime\to\eta X$, as well as inclusive decays of $B$ mesons into final states containing strange hadronic states, $b\to s X$, where the light meson spectra were obtained using \texttt{EPOS-LHC}, while the $b$-quark spectrum was obtained using \texttt{Pythia~8}~\cite{Sjostrand:2006za, Sjostrand:2014zea}. We implement the kaon and $B$-meson branching fractions following Ref.~\cite{Feng:2017vli}, while for $\eta$ mesons we employ the results from Ref.~\cite{Egana-Ugrinovic:2019wzj}. We give the relevant expressions for the branching fractions in \cref{app:BRs}.

The dominant contribution to the $K$ and $B$ decay widths into $X$ corresponds to a top-$W$ loop. Unlike for kaon decays, the relevant branching fractions of $B$ mesons do not suffer from a strong CKM suppression, $V^2_{tb}\gg V_{td}^2$, which makes this channel the dominant production mode despite the suppressed $b$-quark production rate and the broader $p_T$ spectrum. Instead, the decays of $\eta$ mesons are suppressed for small Yukawa-like couplings of $X$ to the first-generation quarks and for a loop-induced coupling to gluons. Similarly to $K_S$, however, prompt decays of $\eta$ mesons make them a non-negligible production mode of $X$. 

\item[Scalar bremsstrahlung] Another production mode of light and high-energy scalar BSM particles produced in the forward direction of the LHC is through their bremsstrahlung in proton-proton collisions. We study this process following the discussion in Ref.~\cite{Boiarska:2019jym} and find that the relevant contribution to the event rate in FASER is typically of order a few $\%$ or smaller.

\item [Lepton-induced production] We have also analyzed a number of production modes that depend on the lepton coupling, $\epsilon_\ell\sim \mathcal{O}(1)$. These include loop-induced Primakoff production from high-energy photons hitting the TAN, cf. Ref.~\cite{Feng:2018noy}, as well as various secondary processes employing electrons and muons pair-produced in the absorber material or traversing the rock shielding of FASER. However, since the difference in $\epsilon_\ell$ and $\epsilon_q$ is typically not sufficient to compensate for a large Yukawa suppression of the corresponding coupling constants, as well as due to additional suppression factors relevant for these production modes, we have found that they play a subdominant role for high-energy scalars of our interest.
\end{description}

In the right panel of \cref{fig:genericandspecificX}, we show the contours with the different number of events in FASER, $N=10,10^2\ldots$, that correspond to the total integrated luminosity of $\mathcal{L}=150~\ifb$ for the LHC Run 3. In the white region of the plot, the KOTO anomaly can be explained without violating current bounds shown as the gray-shaded region, following Ref.~\cite{Liu:2020qgx}. In addition, we have added the constraint from the NuCal experiment following Ref.~\cite{Egana-Ugrinovic:2019wzj}. The green-shaded band in the plot indicates the region in the parameter space of the model in which the $(g-2)_\mu$ anomaly can be resolved. As can be seen, FASER can detect up to $\mathcal{O}(10^4)$ LLP decays in the region of the parameter space corresponding to both the anomalies. In this case, even a few weeks of the operation of the experiment would be enough to test such scenarios. Once more data are accumulated, the entire allowed region of the parameter space will be covered with at least a few hundred expected events.

The breakdown of the number of the expected events in FASER and FASER~2 for different production modes of $X$ is shown in \cref{tab:productionmodes} for the benchmark scenario with $m_X=50~\mev$, $\epsilon_q = 0.016$ and $\epsilon_\ell = 1.22$. The values of the parameters of the model have been chosen so that both the KOTO and $(g-2)_\mu$ anomalies can be fitted to their central values~\cite{Liu:2020qgx}. The dominant production mode in FASER, in this case, is due to rare $B$ meson decays, although decays of lighter mesons give a contribution of order $30\%$ in total. This is dominated by rare decays of $\eta$ mesons. Instead, the kaon decays give only a few $\%$ contribution.

\subsection{\label{sec:longlifeX}A very long-lived dark scalar}

If a new LLP is produced in rare kaon decays with the mass close to the pion mass, $m_X\sim m_\pi$, and the relevant branching fraction fitting the anomaly, cf. \cref{eq:K0Lneutanomaly}, the stringent NA62 and E949 bounds on $K^+$ decays can be avoided even for a very long-lived or effectively stable $X$. This is due to increased BG in the respective searches for $K^+$ decays (see the discussion in Ref.~\cite{Kitahara:2019lws} and references therein). In addition, such scenarios can also escape constraints from beam-dump searches provided that $\tau_X$ is large enough and the LLPs typically overshoot the detector.

An interesting example of such a scenario that can fit the KOTO anomaly is a dark Higgs boson $X$ with a universal mixing angle with the SM species, cf. \cref{eq:L2HDM} with $\epsilon_\ell = \epsilon_q = \epsilon_W$ and Ref.~\cite{Egana-Ugrinovic:2019wzj} for further discussion. In this scenario, the lifetime of the LLP is typically large, $c\tau_X\sim 100~\km$. As a result, probing this model goes beyond the capabilities of FASER and FASER~2 experiments, which are designed to focus on more short-lived BSM species. 

On the other hand, in the long-lifetime regime relevant here, the sensitivity reach in searches for displaced decays of $X$ can be improved by increasing the size of the detector and its angular coverage. We illustrate this in the left panel of \cref{fig:darkHiggsandX1X2}, where we show the expected reach of proposed future Codex-b~\cite{Gligorov:2017nwh}, MATHUSLA~\cite{Curtin:2018mvb} and SHiP~\cite{Alekhin:2015byh} experiments in their searches for dark Higgs boson decays into electron-positron pairs, $X\to e^+e^-$, following Ref.~\cite{Beacham:2019nyx}. 

For comparison, we also present the expected sensitivity of the enlarged FASER~2 detector assuming the total integrated luminosity of $\mathcal{L} = 3~\iab$ relevant for HL-LHC. As can be seen, probing the region in the parameter space of the model that corresponds to the KOTO anomaly would require a quite substantial increase in the detector radius, $R\gtrsim 5~\m$, or its length, $\Delta\gtrsim 50~\m$, with respect to the design discussed in \cref{sec:FASER}. On the other hand, for lower values of $\tau_X$, even a much smaller and properly placed detector can have very good detection prospects, as we show for other scenarios discussed in this study.

\begin{figure*}[tb]
\centering
\includegraphics[width=0.49\textwidth]{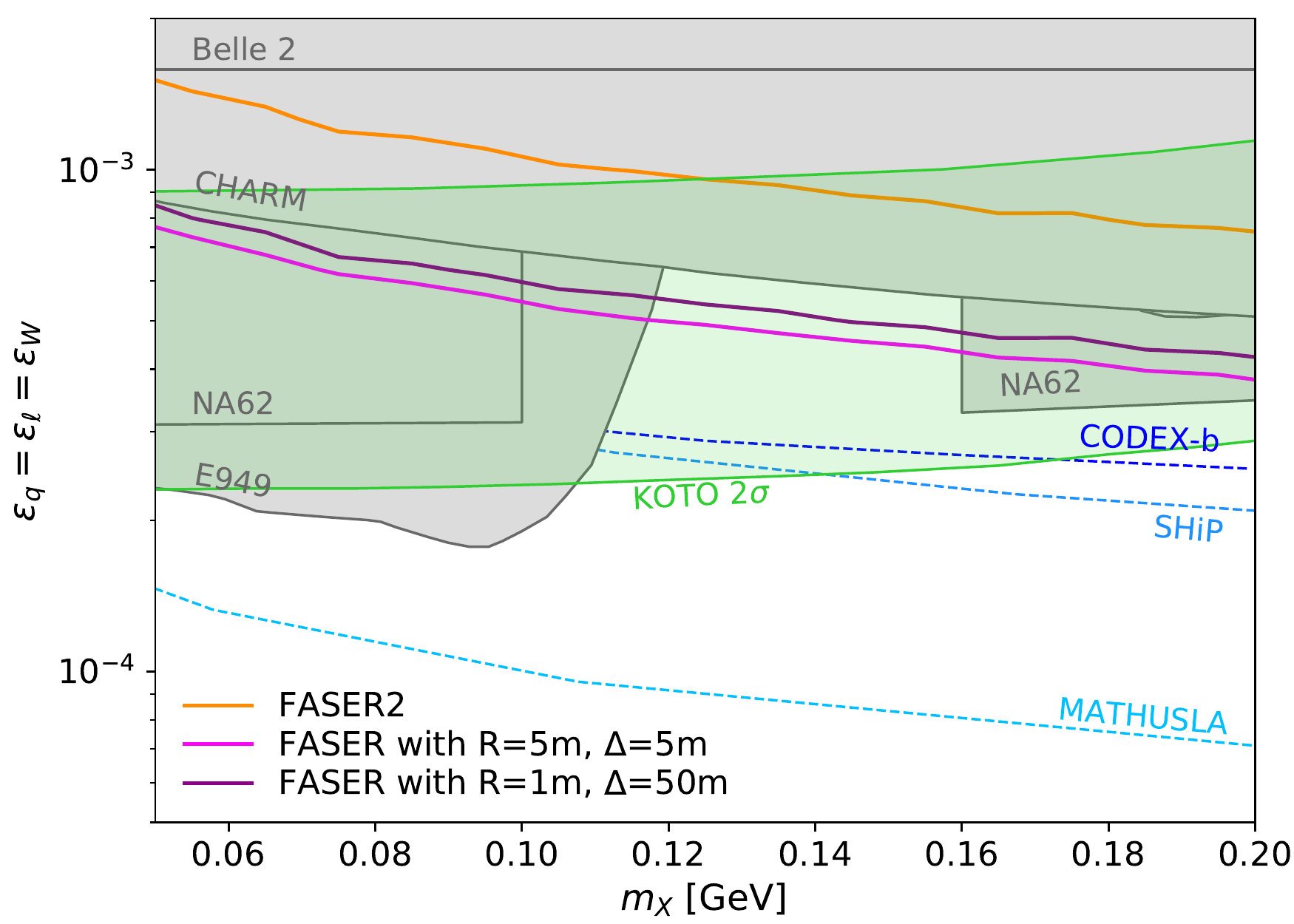}
\hfill
\includegraphics[width=0.49\textwidth]{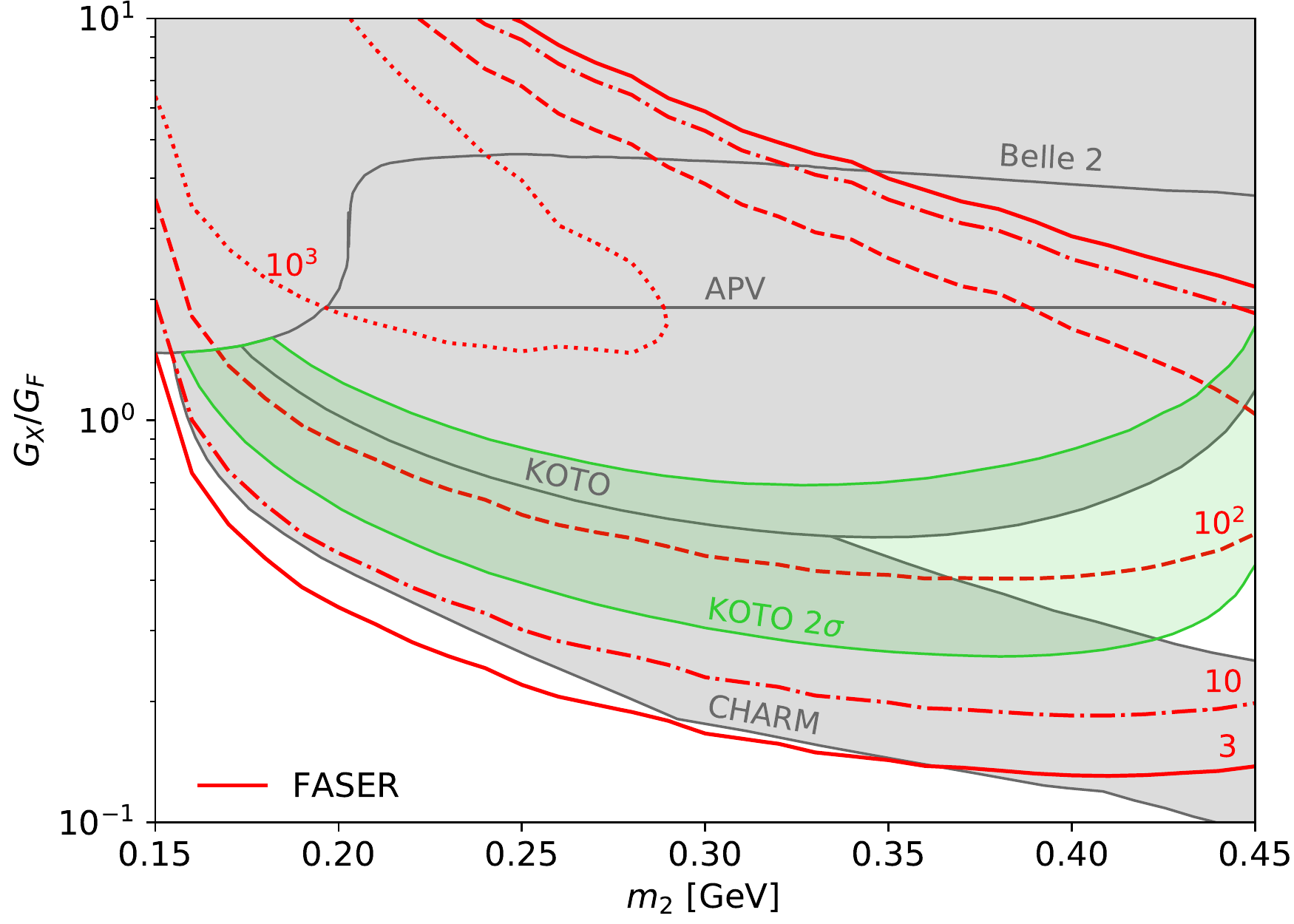}
\caption{The results for the models discussed in \cref{sec:longlifeX} and \cref{sec:X1X2}. In both panels, the $2\sigma$ regions of the parameter space in which the KOTO anomaly can be explained are shown as the green-shaded bands. \textbf{Left:} Reach for the dark Higgs boson model discussed in \cref{sec:longlifeX}. The sensitivity contour for FASER~2 is shown with the orange solid line, while purple and yellow solid lines correspond to larger versions on FASER~2 with the radius $R=5~\m$ or length $\Delta=50~\m$, respectively. The current bounds are shown as gray-shaded regions following Ref.~\cite{Egana-Ugrinovic:2019wzj}. The blue dashed lines correspond to the expected sensitivity of other proposed future experiments (from top to bottom): CODEX-b, SHiP and MATHUSLA. 
\textbf{Right:} Event contours (from outside to inside: 3 (solid red line), 10 (dash-dotted), 100 (dashed), 1000 (dotted)) at FASER at LHC Run 3 for the model with pure dark sector decays of kaons, $K_L\to X_1X_2$, discussed in \cref{sec:X1X2}. The current bounds are shown as gray-shaded regions following Ref.~\cite{Hostert:2020gou}.
}
\label{fig:darkHiggsandX1X2}
\end{figure*}

\subsection{\label{sec:X1X2} $K_L$ decays into dark sector particles}

A different approach to the BSM explanation of the KOTO anomaly has been proposed in Refs~\cite{Gori:2020xvq,Hostert:2020gou}; this approach employs two-body decays of neutral kaons into dark species, $K_L\to \psi_1 \psi_2$. The relevant three-body decay processes of charged kaons, e.g. $K^+\to \pi^+ \psi_1 \psi_2$, can then be kinematically suppressed or even forbidden depending on the masses of the $\psi_i$s. The anomaly can be fitted if at least one of the dark species is unstable and can mimic the di-photon decay signature of neutral pions inside the KOTO detector. 

In the following, we focus on the model in which a non-diagonal coupling of kaons to two dark fermionic states $\psi_i$ arises due to a vector portal and a new gauge field $X_\mu$ that mixes with the SM $Z$ boson~\cite{Babu:1997st, Davoudiasl:2012ag, Dror:2018wfl}
\begin{equation}
\mathcal{L}\supset \frac{1}{2}m_X^2 X_\mu X^\mu - \epsilon_Zm_Z^2X_\mu Z^\mu,
\label{eq:LX1X2}
\end{equation}
where $\epsilon_Z\sim 10^{-3}-10^{-2}$ denotes the relevant mixing parameter. After the electroweak symmetry breaking, a new $Z^\prime$ gauge boson acquires a mass $m^2_{Z^\prime} = m_X^2 - \epsilon_Z^2 m_Z^2$. The interactions of $Z^\prime$ with the dark fermions can then be described by
\begin{equation}
\mathcal{L}\supset g_XZ^\prime_\mu\,(c_V\bar{\psi}_2\gamma^\mu\psi_1 + c_A\bar{\psi}_2\gamma^5\gamma^\mu\psi_1) + h.c.
\end{equation}
In particular, an effective operator which couples $\psi_i$s to $s$ and $d$ quarks is induced at a loop level with $W$ boson exchange~\cite{Dror:2018wfl}. In the following, we set  $g_X=1$, $c_V=0$, $c_A=1$, $m_{Z^\prime} = 10~\gev$, and $m_{2} = 11\,m_{1}$, where $m_i\equiv m_{\psi_i}$.  

The $\psi_2$ decay width and branching fractions for $\psi_1\psi_2$-pair production in various meson decays can be found in Ref.~\cite{Hostert:2020gou}. In particular, as discussed therein, the assumption about a relatively large mass splitting between the dark species, $m_{1}\ll m_{2}-m_{\pi^0}$, allows one to obtain about an $80\%$ branching fraction of two-body decays $\psi_2\to\psi_1\pi^0$. The subsequent prompt decays of high-energy neutral pions with $E_{\pi^0}\gtrsim 100~\gev$ generate a di-photon signature inside the FASER detector.

In the right panel of \cref{fig:darkHiggsandX1X2}, we show the contours with the expected number of events in FASER in the parameter space spanned by $m_{2}$ and $G_X/G_F$ parameters, where $G_F$ is the Fermi coupling constant and $G_X = (\epsilon_Z g g_X)/(2\sqrt{2}c_Wm_{Z^\prime}^2)$ with $g$ corresponding to the weak coupling constant. The region of the parameter space in which a $2\sigma$ fit can be obtained to the KOTO anomaly is indicated by a green-shaded band in the plot. The current bounds on the model from the CHARM experiment, the previous KOTO search for $K_L\to \pi^0\nu\bar{\nu}$, cf. \cref{eq:K0Lneutbound}, atomic parity violation (APV)~\cite{Dror:2018wfl}, and the Belle-II search for $B\to K\nu\bar{\nu}$~\cite{Grygier:2017tzo} are shown as the gray-shaded region following Ref.~\cite{Hostert:2020gou}.\footnote{As discussed in Ref.~\cite{Hostert:2020gou}, a more detailed simulation of the CHARM and NuCal experiments is needed to refine the lower part of the gray-shaded region in \cref{fig:darkHiggsandX1X2}. This, however, is expected to have a small impact on the upper part of the plot corresponding to the KOTO anomaly.}

The region of the parameter space of the model in which the KOTO anomaly can be explained by rare decays $K_L\to \psi_1\psi_2$ corresponds to $350~\mev\lesssim m_{2}\lesssim 450~\mev$. Here, one requires that $\psi_2$ decays within the KOTO detector volume so that it can mimic neutral pion decays. However, such decays that are too fast lead to larger production rates that are already excluded by previous KOTO studies. As a result, the anomalous events can be best fitted for $\tau_{\psi_2}\sim (0.01-0.1)$~ns and such scenarios can be well tested in FASER. In particular, as shown in \cref{fig:darkHiggsandX1X2}, one expects $\mathcal{O}(100)$ of high-energy visible $\psi_2$ decay events in the detector for LHC Run 3. 

\section{\label{sec:ALPs} Beyond the anomaly -- ALP coupling to SU(2)$\mathbf{_W}$ gauge bosons}

The complementarity between FASER and kaon factories extends beyond possible explanations to the currently observed anomalous events in the KOTO experiment. In particular, more general BSM scenarios leading to two-body kaon decays, $K\to\pi^0 X$ and $K^+\to\pi^+ X$, can be constrained in searches for detector-stable $X$ acting as neutrino impostors, as well as in studies focused on displaced decays $X\to\gamma\gamma$ leading to $4\gamma$ or $\pi^+\!+2\gamma$ signatures observed in the detector. Some examples of such studies have recently been discussed in Ref.~\cite{Gori:2020xvq} for models with axion-like particles (ALPs), $X\equiv a$, which couple dominantly to gluons or $SU(2)_W$ gauge bosons. The sensitivity reach of FASER in the former scenario has been studied in Ref.~\cite{Ariga:2018uku}. 

In the following, we will focus on the latter model in which the ALP couples to the SM field strength tensor $W_{\mu\nu}^a$ of the $SU(2)$ group
\begin{equation}
\mathcal{L}\supset -\frac{1}{2}m_a^2 a^2 - \frac{g_{aWW}}{4}a W_{\mu\nu}^a \widetilde{W}_{\mu\nu}^a.
\label{eq:LALP}
\end{equation}
The coupling to $W$ bosons in \cref{eq:LALP} gives rise to both kaon decays, $s\!\to\! d\, a$, and $B$-meson decays, $b\!\to\! s \,a$, via the usual loop diagrams. We implement the relevant branching fractions following Ref.~\cite{Izaguirre:2016dfi,Aloni:2018vki,Gori:2020xvq} as discussed in \cref{app:BRs}. 

After the EWSB, additional couplings of $a$ to $\gamma\gamma$, $Z\gamma$, and $ZZ$ arise with the relative strength dictated by the weak mixing angle. As a result, the ALP can also be produced through the Primakoff process, $\gamma N\to a\gamma$, employing the di-photon coupling $g_{a\gamma\gamma} = g_{aWW}s_W^2$, cf. Ref.~\cite{Feng:2018noy} for relevant discussion for FASER, providing a subleading contribution to the ALP event rate. In contrast, $a$ production in rare $Z$ decays is more isotropic and can be neglected. For the low-mass ALPs of our interest, the dominant decay mode is into two photons, $a \to \gamma\gamma $~\cite{Gavela:2019wzg}. 

We present the FASER, FASER~HL, and FASER~2 sensitivity reach lines for this model in \cref{fig:ALPs}. Additional dashed contours in the plot correspond to the expected future sensitivity in searches in KOTO and NA62 for the aforementioned two-body BSM kaon decays, $K^{(+)}\to \pi^{0(+)}a$. Depending on whether the ALP escapes the detector without decaying or decays inside to two photons, the searches focus on $2\gamma$ or $4\gamma$ signatures in KOTO, and $\pi^++0\gamma$ or $\pi^++2\gamma$ signal in NA62. We show these searches following Ref.~\cite{Gori:2020xvq}. The current bounds on the model are taken from Refs.~\cite{Dolan:2017osp,Gori:2020xvq}. The most important constraints in the region of the parameter space of our interest come from the beam-dump experiment E137 and past searches in kaon factories. 

As can be seen, FASER will cover the entire region of the parameter space corresponding to KOTO and NA62 searches for detector-stable $a$. It will also independently probe scenarios relevant for unstable ALP decaying inside these detectors. In addition, FASER reach extends towards lower values of the $g_{aWW}$ coupling and larger mass of the ALP. The larger values of the coupling constant will, instead, be probed by e.g. Belle-II searches for the $3\gamma$ signature from the ALP production and subsequent prompt decay inside the detector, $e^+e^-\to\gamma+(a\to2\gamma)$. We show the relevant reach for $50~\ifb$ in \cref{fig:ALPs} following Ref.~\cite{Dolan:2017osp}. Similarly, future ATLAS and CMS searches for $Z\to\gamma+(a\to2\gamma)$ signature can also cover an important part of the parameter space of the model corresponding to $g_{aWW}\gtrsim 10^{-5}\,\textrm{GeV}^{-1}$, as shown in the plot based on Ref.~\cite{Bauer:2017ris}.

Both FASER~HL and FASER~2 experiments will be able to further improve the detection prospects, as well as constrain scenarios with $m_a$ up to $1~\gev$ or even close to the limit $m_a\lesssim m_B-m_K$, respectively. In a similar time frame, additional bounds on the model can come from e.g. the SHiP experiment (not shown in the plot).

\begin{figure}[tb]
\centering
\includegraphics[width=0.49\textwidth]{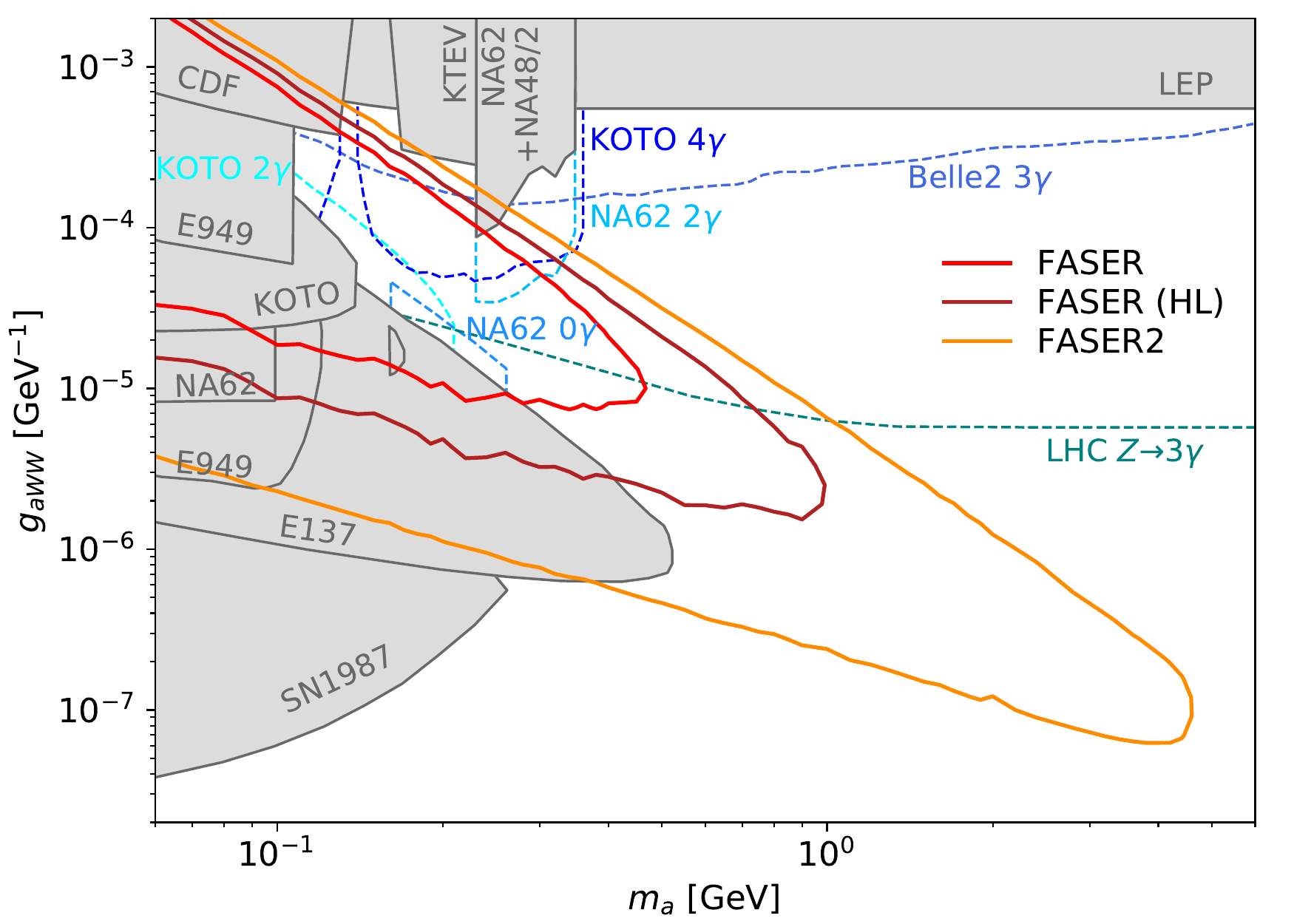}
\caption{Sensitivity reach of the FASER (red solid line), FASER~HL (brown) and FASER~2 (orange) experiments in the model with axion-like particles coupled to $SU(2)_W$ discussed in \cref{sec:ALPs}. The currently excluded gray-shaded region and the dashed blue lines corresponding to the future sensitivity reach in the KOTO and NA62 searches for promptly decaying and detector-stable ALPs are taken from Ref.~\cite{Dolan:2017osp,Gori:2020xvq}. The expected reach of the Belle-II search for the $3\gamma$ signature is shown following Ref.~\cite{Dolan:2017osp}. The future ATLAS/CMS reach in the $Z\to 3\gamma$ signature from production and prompt decays of the ALP to a di-photon pair is shown following Ref.~\cite{Bauer:2017ris}.
} 
\label{fig:ALPs}
\end{figure}

\section{\label{sec:conclusion} Conclusion and Outlook}

A long-awaited definite experimental discovery of new physics effects should first manifest itself as a clear deviation from the SM predictions and excess over expected BG. In this study, we have focused on recently observed and intriguing anomalous neutral kaon decay events in the KOTO experiment. The current KOTO observation corresponds to a striking excess over the expected SM branching fraction of the decay process $K_L\to\pi^0\nu\bar{\nu}$, while it awaits further dedicated analyses to be performed by the KOTO collaboration and other experiments before it can be fully confirmed. In particular, we have analyzed the prospects of independently probing the relevant BSM scenarios proposed to explain the anomaly in the upcoming FASER experiment at the LHC~\cite{Feng:2017uoz,Ariga:2018zuc,Ariga:2018pin}. 

We have shown that FASER could start probing such models, which predict light and unstable new particles, immediately after the beginning of its operation. The $pp$ collisions at the LHC produce large numbers of kaons that typically hit elements of the infrastructure before decaying unless they go down the beam pipe i.e. towards FASER. This can lead to a large flux of forward-going LLPs produced in rare kaon decays and in other production modes. The FASER baseline compatible with the predicted lifetime $c\tau_X$ of such LLPs, and its excellent di-photon detection capabilities, will then result in up to $\mathcal{O}(10^4)$ high-energy LLP decay events observed during LHC Run 3. 

Our results are presented for selected distinct BSM scenarios corresponding to the \textsl{lifetime solution} to the KOTO anomaly, as well as for some other models that either predict very long-lived LLPs with a mass close to the pion mass or employ purely dark sector decays of neutral kaons. Interestingly, even in the less promising case of large $\tau_X$, FASER can also be indirectly sensitive to KOTO-related models that predict a larger set of LLPs and simultaneously address a number of outstanding experimental and theoretical issues. An example of such a scenario with below $\gev$-scale heavy neutral leptons within the reach of FASER 2~\cite{Kling:2018wct,Ariga:2018uku} has recently been discussed in Ref.~\cite{Cline:2020mdt}. Although in this study we focus on collider searches for new physics, it is important to note that the KOTO anomaly can also have profound cosmological consequences that would have to be thoroughly investigated, especially if the observed excess persists, cf. Ref.~\cite{Altmannshofer:2020pjb} for a recent discussion.

Last but not least, FASER complementarity to BSM searches in kaon factories extends beyond the current anomaly. We have illustrated this for a simplified model with ALPs coupled to $SU(2)_W$ gauge bosons that dominantly decay to $\gamma\gamma$ pairs. FASER will probe regions of the parameter space relevant for several distinct searches for LLPs decaying either inside or outside the KOTO and NA62 detectors. It will, therefore, provide an independent probe of such scenarios. In a longer term, improved bounds can be obtained by FASER~HL or FASER~2 detectors operating during the HL-LHC phase.

The upcoming run of the LHC will push the limits of intensity-frontier exploration of new physics to the next level. The FASER experiment will play a vital role in this endeavor toward potentially groundbreaking discoveries.


\acknowledgments
We would like to thank Jonathan Feng, Sam Homiller and Ben Lillard for useful discussions and comments on the manuscript. We are also grateful to the authors and maintainers of many open-source software packages, including
\texttt{CRMC}~\cite{CRMC},
\texttt{EPOS}~\cite{Pierog:2013ria}
\texttt{Jupyter} notebooks~\cite{soton403913}, 
\texttt{Matplotlib}~\cite{Hunter:2007}, 
\texttt{NumPy}~\cite{numpy}, 
\texttt{pylhe}~\cite{lukas_2018_1217032}, 
\texttt{Pythia~8}~\cite{Sjostrand:2014zea},
and \texttt{scikit-hep}~\cite{Rodrigues:2019nct}.
FK is supported by the Department of Energy under Grant DE-AC02-76SF00515. ST is supported by the Lancaster-Manchester-Sheffield Consortium for Fundamental Physics under STFC grant ST/P000800/1. ST is partially supported by the Polish Ministry of Science and Higher Education through its scholarship for young and outstanding scientists (decision no. 1190/E-78/STYP/14/2019).

\appendix

\section{\label{app:BRs} Meson decay branching fractions to dark Higgs boson and ALP}

Below, we list the branching fractions of the dominant meson decay modes to produce light scalar and pseudoscalar particles in the models discussed in \cref{sec:2HDM}, \cref{sec:longlifeX} and \cref{sec:ALPs}.

\begin{description}
\item[Scalar decay] For a light scalar, as defined in \cref{eq:L2HDM}, the dominant production modes are rare decays of $\eta$s, $\eta'$s, kaons and $b$-quarks, with the following branching fractions adapted from Refs.~\cite{Feng:2017vli, Egana-Ugrinovic:2019wzj} and references therein
\be
\mathcal{B}(K^+ \!\to\! \pi^+ X) &\simeq 2.0 \cdot 10^{-3} \times \epsilon_q^2\, \lambda^{\frac 1 2}(m_K,m_\pi,m_a), \!\!\\
\mathcal{B}(K_L \!\to\! \pi^0 X) &\simeq 7.0 \cdot 10^{-3} \times \epsilon_q^2 \, \lambda^{\frac 1 2}(m_K,m_\pi,m_a), \!\!\\
\mathcal{B}(K_S \!\to\! \pi^0 X) &\simeq 2.2 \cdot 10^{-6} \times \epsilon_q^2 \, \lambda^{\frac 1 2}(m_K,m_\pi,m_a), \!\!\\
\mathcal{B}(\eta \!\to\! \pi^0 X) &\simeq 3.4 \cdot 10^{-5} \times \epsilon_q^2 \, \lambda^{\frac 1 2}(m_\eta,m_\pi,m_a), \!\!\\
\mathcal{B}(\eta' \!\to\! \eta X) &\simeq 7.2 \cdot 10^{-5} \times \epsilon_q^2 \, \lambda^{\frac 1 2}(m_{\eta'},m_\eta,m_a), \!\!\\
\mathcal{B}(b \!\to\! s X) &\simeq 5.7 \times \epsilon_q^2 \,\left(1-m_X^2/m_b^2\right)^2, \nonumber
\ee
with 
\be
\lambda(m_1,m_2,m_3)\!=\!\left[1\!-\!\frac{(m_2\!+\!m_3)^2}{m_1^2} \right]\! \left[1\!-\!\frac{(m_2\!-\!m_3)^2}{m_1^2} \right] \! . \!\!\!\!\!\! \nonumber
\ee

\item[ALP with the $\mathbf{SU(2)_W}$ couplings] For an ALP, as defined in \cref{eq:LALP} and with $g \!=\! g_{aWW} \cdot \gev$, the relevant branching fractions for kaon decays read~\cite{Gori:2020xvq},
\be
\mathcal{B}(K^+ \!\to\! \pi^+ a) &\simeq 10.5\times g^2
\lambda^{\frac 1 2}(m_K,m_\pi,m_a), \!\!\!\!\\
\mathcal{B}(K_L \!\to\! \pi^0 a) &\simeq 4.5\times g^2
\lambda^{\frac 1 2}(m_K,m_\pi,m_a). \!\!\!\! 
\nonumber
\ee
In the case of B-mesons, we follow the data-driven approach discussed in Ref.~\cite{Aloni:2018vki} and use $\mathcal{B}(b\!\to\!sa) \approx 5 \times  [\mathcal{B}(B\!\to\! Ka) \!+\! \mathcal{B}(B \!\to\! K^{*} a)]$ with the individual branching fractions taken from  Ref.~\cite{Izaguirre:2016dfi} 
\be
\mathcal{B}(B \!\to\! K a) &\simeq 2\cdot 10^4 \times g^2
\lambda^{\frac 1 2}(m_B,m_K,m_a)  F_{K}^2,  \!\!\!\!\\
\mathcal{B}(B \!\to\! K^* a) &\simeq 2\cdot 10^4 \times  g^2
\lambda^{\frac 3 2}(m_B,m_{K^*},m_a) F_{K^*}^2,  \!\!\!\!\nonumber
\ee
where the relevant form factors are given by~\cite{Ball:2004ye,Ball:2004rg}
\be
F_K = \frac{0.33}{1-m_a^2/(38~\gev^2)} 
\!\!\!\!\nonumber
\ee
and
\be
F_{K^*} = \frac{1.35}{1-m_a^2/(28~\gev^2)} - \frac{1}{1-m_a^2/(37~\gev^2)} .
\!\!\!\! \nonumber
\ee
\end{description}


\bibliography{references}

\providecommand{\href}[2]{#2}\begingroup\raggedright\begin{thebibliography}{10}

\bibitem{KOTOanomaly}
S.~Shinohara, ``Search for the rare decay $\uppercase{K}_{\uppercase{l}} \to
  \pi^0\nu\overline{\nu}$ at \uppercase{J-PARC KOTO} experiment.'' KAON 2019,
  2019.

\bibitem{Kitahara:2019lws}
T.~Kitahara, T.~Okui, G.~Perez, Y.~Soreq, and K.~Tobioka, ``{New physics
  implications of recent search for $K_L \to \pi^0 \nu\bar{\nu}$ at KOTO},''
  \href{http://dx.doi.org/10.1103/PhysRevLett.124.071801}{{\em Phys. Rev.
  Lett.} {\bf 124} (2020) no.~7, 071801},
  \href{http://arxiv.org/abs/1909.11111}{{\tt arXiv:1909.11111 [hep-ph]}}.

\bibitem{Feng:2017uoz}
J.~L. Feng, I.~Galon, F.~Kling, and S.~Trojanowski, ``{ForwArd Search
  ExpeRiment at the LHC},''
  \href{http://dx.doi.org/10.1103/PhysRevD.97.035001}{{\em Phys. Rev. D} {\bf
  97} (2018) no.~3, 035001}, \href{http://arxiv.org/abs/1708.09389}{{\tt
  arXiv:1708.09389 [hep-ph]}}.

\bibitem{Ariga:2018zuc}
{\bf FASER} Collaboration, A.~Ariga {\em et al.}, ``{Letter of Intent for
  FASER: ForwArd Search ExpeRiment at the LHC},''
  \href{http://arxiv.org/abs/1811.10243}{{\tt arXiv:1811.10243
  [physics.ins-det]}}. \url{https://cds.cern.ch/record/2642351}.
Submitted to the CERN LHCC on 18 July 2018.

\bibitem{Ariga:2018pin}
{\bf FASER} Collaboration, A.~Ariga {\em et al.}, ``{Technical Proposal for
  FASER: ForwArd Search ExpeRiment at the LHC},''
  \href{http://arxiv.org/abs/1812.09139}{{\tt arXiv:1812.09139
  [physics.ins-det]}}. \url{http://cds.cern.ch/record/2651328}.
Submitted to the CERN LHCC on 7 November 2018.

\bibitem{Yamanaka:2012yma}
{\bf KOTO} Collaboration, T.~Yamanaka, ``{The J-PARC KOTO experiment},''
  \href{http://dx.doi.org/10.1093/ptep/pts057}{{\em PTEP} {\bf 2012} (2012)
  02B006}.

\bibitem{Nagamiya:2012tma}
S.~Nagamiya, ``{Introduction to J-PARC},''
  \href{http://dx.doi.org/10.1093/ptep/pts025}{{\em PTEP} {\bf 2012} (2012)
  02B001}.

\bibitem{Ahn:2018mvc}
{\bf KOTO} Collaboration, J.~Ahn {\em et al.}, ``{Search for the $K_L \!\to\!
  \pi^0 \nu \overline{\nu}$ and $K_L \!\to\! \pi^0 X^0$ decays at the J-PARC
  KOTO experiment},''
  \href{http://dx.doi.org/10.1103/PhysRevLett.122.021802}{{\em Phys. Rev.
  Lett.} {\bf 122} (2019) no.~2, 021802},
  \href{http://arxiv.org/abs/1810.09655}{{\tt arXiv:1810.09655 [hep-ex]}}.

\bibitem{Buras:2015qea}
A.~J. Buras, D.~Buttazzo, J.~Girrbach-Noe, and R.~Knegjens, ``{$ {K}^{+}\to
  {\pi}^{+}\nu \overline{\nu} $ and $ {K}_L\to {\pi}^0\nu \overline{\nu} $ in
  the Standard Model: status and perspectives},''
  \href{http://dx.doi.org/10.1007/JHEP11(2015)033}{{\em JHEP} {\bf 11} (2015)
  033}, \href{http://arxiv.org/abs/1503.02693}{{\tt arXiv:1503.02693
  [hep-ph]}}.

\bibitem{Grossman:1997sk}
Y.~Grossman and Y.~Nir, ``{$K_L \to \pi^0\nu\bar{\nu}$ beyond the standard
  model},'' \href{http://dx.doi.org/10.1016/S0370-2693(97)00210-4}{{\em Phys.
  Lett. B} {\bf 398} (1997)  163--168},
  \href{http://arxiv.org/abs/hep-ph/9701313}{{\tt arXiv:hep-ph/9701313}}.

\bibitem{Artamonov:2008qb}
{\bf E949} Collaboration, A.~Artamonov {\em et al.}, ``{New measurement of the
  $K^{+} \to \pi^{+} \nu \bar{\nu}$ branching ratio},''
  \href{http://dx.doi.org/10.1103/PhysRevLett.101.191802}{{\em Phys. Rev.
  Lett.} {\bf 101} (2008)  191802}, \href{http://arxiv.org/abs/0808.2459}{{\tt
  arXiv:0808.2459 [hep-ex]}}.

\bibitem{Artamonov:2009sz}
{\bf BNL-E949} Collaboration, A.~Artamonov {\em et al.}, ``{Study of the decay
  $K^+\to\pi^+\nu \bar\nu$ in the momentum region $140 < P_\pi < 199$ MeV/c},''
  \href{http://dx.doi.org/10.1103/PhysRevD.79.092004}{{\em Phys. Rev. D} {\bf
  79} (2009)  092004}, \href{http://arxiv.org/abs/0903.0030}{{\tt
  arXiv:0903.0030 [hep-ex]}}.

\bibitem{NA62bound}
G.~Ruggiero, ``New result on $\uppercase{K}^+\to\pi^+\nu\overline{\nu}$ from
  the \uppercase{NA62} experiment.'' KAON2019, 2019.

\bibitem{Fabbrichesi:2019bmo}
M.~Fabbrichesi and E.~Gabrielli, ``{Dark-sector physics in the search for the
  rare decays $K^+\rightarrow \pi^+ \bar \nu \nu$ and $K_L\rightarrow \pi^0
  \bar \nu \nu$},'' \href{http://arxiv.org/abs/1911.03755}{{\tt
  arXiv:1911.03755 [hep-ph]}}.

\bibitem{Egana-Ugrinovic:2019wzj}
D.~Egana-Ugrinovic, S.~Homiller, and P.~Meade, ``{Light Scalars and the KOTO
  Anomaly},'' \href{http://dx.doi.org/10.1103/PhysRevLett.124.191801}{{\em
  Phys. Rev. Lett.} {\bf 124} (2020) no.~19, 191801},
  \href{http://arxiv.org/abs/1911.10203}{{\tt arXiv:1911.10203 [hep-ph]}}.

\bibitem{Li:2019fhz}
T.~Li, X.-D. Ma, and M.~A. Schmidt, ``{Implication of $K\to \pi \nu \bar{\nu}$
  for generic neutrino interactions in effective field theories},''
  \href{http://dx.doi.org/10.1103/PhysRevD.101.055019}{{\em Phys. Rev. D} {\bf
  101} (2020) no.~5, 055019}, \href{http://arxiv.org/abs/1912.10433}{{\tt
  arXiv:1912.10433 [hep-ph]}}.

\bibitem{Dev:2019hho}
P.~B. Dev, R.~N. Mohapatra, and Y.~Zhang, ``{Constraints on long-lived light
  scalars with flavor-changing couplings and the KOTO anomaly},''
  \href{http://dx.doi.org/10.1103/PhysRevD.101.075014}{{\em Phys. Rev. D} {\bf
  101} (2020) no.~7, 075014}, \href{http://arxiv.org/abs/1911.12334}{{\tt
  arXiv:1911.12334 [hep-ph]}}.

\bibitem{Liu:2020qgx}
J.~Liu, N.~McGinnis, C.~E. Wagner, and X.-P. Wang, ``{A light scalar
  explanation of $(g-2)_{\mu}$ and the KOTO anomaly},''
  \href{http://dx.doi.org/10.1007/JHEP04(2020)197}{{\em JHEP} {\bf 04} (2020)
  197}, \href{http://arxiv.org/abs/2001.06522}{{\tt arXiv:2001.06522
  [hep-ph]}}.

\bibitem{Jho:2020jsa}
Y.~Jho, S.~M. Lee, S.~C. Park, Y.~Park, and P.-Y. Tseng, ``{Light gauge boson
  interpretation for $(g- 2)_{\mu}$ and the $K_{L}\rightarrow \pi^{0} + $
  (invisible) anomaly at the J-PARC KOTO experiment},''
  \href{http://dx.doi.org/10.1007/JHEP04(2020)086}{{\em JHEP} {\bf 04} (2020)
  086}, \href{http://arxiv.org/abs/2001.06572}{{\tt arXiv:2001.06572
  [hep-ph]}}.

\bibitem{Cline:2020mdt}
J.~M. Cline, M.~Puel, and T.~Toma, ``{A little theory of everything, with heavy
  neutral leptons},'' \href{http://dx.doi.org/10.1007/JHEP05(2020)039}{{\em
  JHEP} {\bf 05} (2020)  039}, \href{http://arxiv.org/abs/2001.11505}{{\tt
  arXiv:2001.11505 [hep-ph]}}.

\bibitem{He:2020jzn}
X.-G. He, X.-D. Ma, J.~Tandean, and G.~Valencia, ``{Breaking the Grossman-Nir
  Bound in Kaon Decays},''
  \href{http://dx.doi.org/10.1007/JHEP04(2020)057}{{\em JHEP} {\bf 04} (2020)
  057}, \href{http://arxiv.org/abs/2002.05467}{{\tt arXiv:2002.05467
  [hep-ph]}}.

\bibitem{Ziegler:2020ize}
R.~Ziegler, J.~Zupan, and R.~Zwicky, ``{Three Exceptions to the Grossman-Nir
  Bound},'' \href{http://arxiv.org/abs/2005.00451}{{\tt arXiv:2005.00451
  [hep-ph]}}.

\bibitem{Liao:2020boe}
Y.~Liao, H.-L. Wang, C.-Y. Yao, and J.~Zhang, ``{An imprint of a new light
  particle at KOTO?},'' \href{http://arxiv.org/abs/2005.00753}{{\tt
  arXiv:2005.00753 [hep-ph]}}.

\bibitem{He:2020jly}
X.-G. He, X.-D. Ma, J.~Tandean, and G.~Valencia, ``{Evading the Grossman-Nir
  bound with $\Delta I=3/2$ new physics},''
  \href{http://arxiv.org/abs/2005.02942}{{\tt arXiv:2005.02942 [hep-ph]}}.

\bibitem{Gori:2020xvq}
S.~Gori, G.~Perez, and K.~Tobioka, ``{KOTO vs. NA62 Dark Scalar Searches},''
  \href{http://arxiv.org/abs/2005.05170}{{\tt arXiv:2005.05170 [hep-ph]}}.

\bibitem{Hostert:2020gou}
M.~Hostert, K.~Kaneta, and M.~Pospelov, ``{Pair production of dark particles in
  meson decays},'' \href{http://arxiv.org/abs/2005.07102}{{\tt arXiv:2005.07102
  [hep-ph]}}.

\bibitem{Datta:2020auq}
A.~Datta, S.~Kamali, and D.~Marfatia, ``{Dark sector origin of the KOTO and
  MiniBooNE anomalies},'' \href{http://arxiv.org/abs/2005.08920}{{\tt
  arXiv:2005.08920 [hep-ph]}}.

\bibitem{Dutta:2020scq}
B.~Dutta, S.~Ghosh, and T.~Li, ``{Explaining $(g-2)_{\mu,e}$, KOTO anomaly and
  MinibooNE excess in an extended Higgs model with sterile neutrinos},''
  \href{http://arxiv.org/abs/2006.01319}{{\tt arXiv:2006.01319 [hep-ph]}}.

\bibitem{Altmannshofer:2020pjb}
W.~Altmannshofer, B.~V. Lehmann, and S.~Profumo, ``{Cosmological implications
  of the KOTO excess},'' \href{http://arxiv.org/abs/2006.05064}{{\tt
  arXiv:2006.05064 [hep-ph]}}.

\bibitem{Liu:2020ser}
X.~Liu, Y.~Li, T.~Li, and B.~Zhu, ``{The Light Sgoldstino Phenomenology:
  Explanations for the Muon $(g-2)$ Deviation and KOTO Anomaly},''
  \href{http://arxiv.org/abs/2006.08869}{{\tt arXiv:2006.08869 [hep-ph]}}.

\bibitem{Archer-Smith:2020hqq}
P.~Archer-Smith and Y.~Zhang, ``{Higgs Portal From The Atmosphere To
  Hyper-K},'' \href{http://arxiv.org/abs/2005.08980}{{\tt arXiv:2005.08980
  [hep-ph]}}.

\bibitem{Fuyuto:2014cya}
K.~Fuyuto, W.-S. Hou, and M.~Kohda, ``{Loophole in $K \to \pi\nu\bar{\nu}$
  Search and New Weak Leptonic Forces},''
  \href{http://dx.doi.org/10.1103/PhysRevLett.114.171802}{{\em Phys. Rev.
  Lett.} {\bf 114} (2015)  171802}, \href{http://arxiv.org/abs/1412.4397}{{\tt
  arXiv:1412.4397 [hep-ph]}}.

\bibitem{Beacham:2019nyx}
J.~Beacham {\em et al.}, ``{Physics Beyond Colliders at CERN: Beyond the
  Standard Model Working Group Report},''
  \href{http://dx.doi.org/10.1088/1361-6471/ab4cd2}{{\em J. Phys. G} {\bf 47}
  (2020) no.~1, 010501}, \href{http://arxiv.org/abs/1901.09966}{{\tt
  arXiv:1901.09966 [hep-ex]}}.

\bibitem{Alimena:2019zri}
J.~Alimena {\em et al.}, ``{Searching for Long-Lived Particles beyond the
  Standard Model at the Large Hadron Collider},''
  \href{http://arxiv.org/abs/1903.04497}{{\tt arXiv:1903.04497 [hep-ex]}}.

\bibitem{Feng:2017vli}
J.~L. Feng, I.~Galon, F.~Kling, and S.~Trojanowski, ``{Dark Higgs bosons at the
  ForwArd Search ExpeRiment},''
  \href{http://dx.doi.org/10.1103/PhysRevD.97.055034}{{\em Phys. Rev.} {\bf
  D97} (2018) no.~5, 055034},
\href{http://arxiv.org/abs/1710.09387}{{\tt arXiv:1710.09387 [hep-ph]}}.

\bibitem{Kling:2018wct}
F.~Kling and S.~Trojanowski, ``{Heavy Neutral Leptons at FASER},''
  \href{http://dx.doi.org/10.1103/PhysRevD.97.095016}{{\em Phys. Rev.} {\bf
  D97} (2018) no.~9, 095016},
\href{http://arxiv.org/abs/1801.08947}{{\tt arXiv:1801.08947 [hep-ph]}}.

\bibitem{Feng:2018noy}
J.~L. Feng, I.~Galon, F.~Kling, and S.~Trojanowski, ``{Axionlike particles at
  FASER: The LHC as a photon beam dump},''
  \href{http://dx.doi.org/10.1103/PhysRevD.98.055021}{{\em Phys. Rev.} {\bf
  D98} (2018) no.~5, 055021},
\href{http://arxiv.org/abs/1806.02348}{{\tt arXiv:1806.02348 [hep-ph]}}.

\bibitem{Ariga:2018uku}
{\bf FASER} Collaboration, A.~Ariga {\em et al.}, ``{FASER’s physics reach
  for long-lived particles},''
  \href{http://dx.doi.org/10.1103/PhysRevD.99.095011}{{\em Phys. Rev.} {\bf
  D99} (2019) no.~9, 095011},
\href{http://arxiv.org/abs/1811.12522}{{\tt arXiv:1811.12522 [hep-ph]}}.

\bibitem{Berlin:2018jbm}
A.~Berlin and F.~Kling, ``{Inelastic Dark Matter at the LHC Lifetime Frontier:
  ATLAS, CMS, LHCb, CODEX-b, FASER, and MATHUSLA},''
  \href{http://dx.doi.org/10.1103/PhysRevD.99.015021}{{\em Phys. Rev.} {\bf
  D99} (2019) no.~1, 015021},
\href{http://arxiv.org/abs/1810.01879}{{\tt arXiv:1810.01879 [hep-ph]}}.

\bibitem{Ariga:2019ufm}
{\bf FASER} Collaboration, A.~Ariga {\em et al.}, ``{FASER: ForwArd Search
  ExpeRiment at the LHC},'' \href{http://arxiv.org/abs/1901.04468}{{\tt
  arXiv:1901.04468 [hep-ex]}}.

\bibitem{Jodlowski:2019ycu}
K.~Jod{\l}owski, F.~Kling, L.~Roszkowski, and S.~Trojanowski, ``{Extending the
  reach of FASER, MATHUSLA, and SHiP towards smaller lifetimes using secondary
  particle production},''
  \href{http://dx.doi.org/10.1103/PhysRevD.101.095020}{{\em Phys. Rev. D} {\bf
  101} (2020) no.~9, 095020}, \href{http://arxiv.org/abs/1911.11346}{{\tt
  arXiv:1911.11346 [hep-ph]}}.

\bibitem{Abreu:2019yak}
{\bf FASER} Collaboration, H.~Abreu {\em et al.}, ``{Detecting and Studying
  High-Energy Collider Neutrinos with FASER at the LHC},''
\href{http://arxiv.org/abs/1908.02310}{{\tt arXiv:1908.02310 [hep-ex]}}.

\bibitem{Abreu:2020ddv}
{\bf FASER} Collaboration, H.~Abreu {\em et al.}, ``{Technical Proposal:
  FASERnu},''
\href{http://arxiv.org/abs/2001.03073}{{\tt arXiv:2001.03073
  [physics.ins-det]}}.

\bibitem{Kling:2020iar}
F.~Kling, ``{Probing Light Gauge Bosons in Tau Neutrino Experiments},''
  \href{http://arxiv.org/abs/2005.03594}{{\tt arXiv:2005.03594 [hep-ph]}}.

\bibitem{Tanabashi:2018oca}
{\bf Particle Data Group} Collaboration, M.~Tanabashi {\em et al.}, ``{Review
  of Particle Physics},''
  \href{http://dx.doi.org/10.1103/PhysRevD.98.030001}{{\em Phys. Rev. D} {\bf
  98} (2018) no.~3, 030001}.

\bibitem{Bergsma:1985qz}
{\bf CHARM} Collaboration, F.~Bergsma {\em et al.}, ``{Search for Axion Like
  Particle Production in 400-{GeV} Proton - Copper Interactions},''
  \href{http://dx.doi.org/10.1016/0370-2693(85)90400-9}{{\em Phys. Lett. B}
  {\bf 157} (1985)  458--462}.

\bibitem{Blumlein:1990ay}
J.~Blumlein {\em et al.}, ``{Limits on neutral light scalar and pseudoscalar
  particles in a proton beam dump experiment},''
  \href{http://dx.doi.org/10.1007/BF01548556}{{\em Z. Phys. C} {\bf 51} (1991)
  341--350}.

\bibitem{Blumlein:1991xh}
J.~Blumlein {\em et al.}, ``{Limits on the mass of light (pseudo)scalar
  particles from Bethe-Heitler e+ e- and mu+ mu- pair production in a proton -
  iron beam dump experiment},''
  \href{http://dx.doi.org/10.1142/S0217751X9200171X}{{\em Int. J. Mod. Phys. A}
  {\bf 7} (1992)  3835--3850}.

\bibitem{Abouzaid:2008xm}
{\bf KTeV} Collaboration, E.~Abouzaid {\em et al.}, ``{Final Results from the
  KTeV Experiment on the Decay $K_{L} \to \pi^0 \gamma \gamma$},''
  \href{http://dx.doi.org/10.1103/PhysRevD.77.112004}{{\em Phys. Rev. D} {\bf
  77} (2008)  112004}, \href{http://arxiv.org/abs/0805.0031}{{\tt
  arXiv:0805.0031 [hep-ex]}}.

\bibitem{Pierog:2013ria}
T.~Pierog, I.~Karpenko, J.~Katzy, E.~Yatsenko, and K.~Werner, ``{EPOS LHC: Test
  of collective hadronization with data measured at the CERN Large Hadron
  Collider},'' \href{http://dx.doi.org/10.1103/PhysRevC.92.034906}{{\em Phys.
  Rev. C} {\bf 92} (2015) no.~3, 034906},
  \href{http://arxiv.org/abs/1306.0121}{{\tt arXiv:1306.0121 [hep-ph]}}.

\bibitem{CRMC}
C.~Baus, T.~Pierog, and R.~Ulrich, ``{Cosmic Ray Monte Carlo (CRMC)},''.
  \url{https://web.ikp.kit.edu/rulrich/crmc.html}.

\bibitem{Alekhin:2015byh}
S.~Alekhin {\em et al.}, ``{A facility to Search for Hidden Particles at the
  CERN SPS: the SHiP physics case},''
  \href{http://dx.doi.org/10.1088/0034-4885/79/12/124201}{{\em Rept. Prog.
  Phys.} {\bf 79} (2016) no.~12, 124201},
  \href{http://arxiv.org/abs/1504.04855}{{\tt arXiv:1504.04855 [hep-ph]}}.

\bibitem{Bennett:2006fi}
{\bf Muon g-2} Collaboration, G.~Bennett {\em et al.}, ``{Final Report of the
  Muon E821 Anomalous Magnetic Moment Measurement at BNL},''
  \href{http://dx.doi.org/10.1103/PhysRevD.73.072003}{{\em Phys. Rev. D} {\bf
  73} (2006)  072003}, \href{http://arxiv.org/abs/hep-ex/0602035}{{\tt
  arXiv:hep-ex/0602035}}.

\bibitem{Bjorken:1988as}
J.~Bjorken, S.~Ecklund, W.~Nelson, A.~Abashian, C.~Church, B.~Lu, L.~Mo,
  T.~Nunamaker, and P.~Rassmann, ``{Search for Neutral Metastable Penetrating
  Particles Produced in the SLAC Beam Dump},''
  \href{http://dx.doi.org/10.1103/PhysRevD.38.3375}{{\em Phys. Rev. D} {\bf 38}
  (1988)  3375}.

\bibitem{Davier:1989wz}
M.~Davier and H.~Nguyen~Ngoc, ``{An Unambiguous Search for a Light Higgs
  Boson},'' \href{http://dx.doi.org/10.1016/0370-2693(89)90174-3}{{\em Phys.
  Lett. B} {\bf 229} (1989)  150--155}.

\bibitem{AlaviHarati:2003mr}
{\bf KTeV} Collaboration, A.~Alavi-Harati {\em et al.}, ``{Search for the rare
  decay K(L) ---> pi0 e+ e-},''
  \href{http://dx.doi.org/10.1103/PhysRevLett.93.021805}{{\em Phys. Rev. Lett.}
  {\bf 93} (2004)  021805}, \href{http://arxiv.org/abs/hep-ex/0309072}{{\tt
  arXiv:hep-ex/0309072}}.

\bibitem{Sjostrand:2006za}
T.~Sjostrand, S.~Mrenna, and P.~Z. Skands, ``{PYTHIA 6.4 Physics and Manual},''
  \href{http://dx.doi.org/10.1088/1126-6708/2006/05/026}{{\em JHEP} {\bf 05}
  (2006)  026}, \href{http://arxiv.org/abs/hep-ph/0603175}{{\tt
  arXiv:hep-ph/0603175}}.

\bibitem{Sjostrand:2014zea}
T.~Sjöstrand, S.~Ask, J.~R. Christiansen, R.~Corke, N.~Desai, P.~Ilten,
  S.~Mrenna, S.~Prestel, C.~O. Rasmussen, and P.~Z. Skands, ``{An Introduction
  to PYTHIA 8.2},'' \href{http://dx.doi.org/10.1016/j.cpc.2015.01.024}{{\em
  Comput. Phys. Commun.} {\bf 191} (2015)  159--177},
  \href{http://arxiv.org/abs/1410.3012}{{\tt arXiv:1410.3012 [hep-ph]}}.

\bibitem{Boiarska:2019jym}
I.~Boiarska, K.~Bondarenko, A.~Boyarsky, V.~Gorkavenko, M.~Ovchynnikov, and
  A.~Sokolenko, ``{Phenomenology of GeV-scale scalar portal},''
  \href{http://dx.doi.org/10.1007/JHEP11(2019)162}{{\em JHEP} {\bf 11} (2019)
  162}, \href{http://arxiv.org/abs/1904.10447}{{\tt arXiv:1904.10447
  [hep-ph]}}.

\bibitem{Gligorov:2017nwh}
V.~V. Gligorov, S.~Knapen, M.~Papucci, and D.~J. Robinson, ``{Searching for
  Long-lived Particles: A Compact Detector for Exotics at LHCb},''
  \href{http://dx.doi.org/10.1103/PhysRevD.97.015023}{{\em Phys. Rev. D} {\bf
  97} (2018) no.~1, 015023}, \href{http://arxiv.org/abs/1708.09395}{{\tt
  arXiv:1708.09395 [hep-ph]}}.

\bibitem{Curtin:2018mvb}
D.~Curtin {\em et al.}, ``{Long-Lived Particles at the Energy Frontier: The
  MATHUSLA Physics Case},''
  \href{http://dx.doi.org/10.1088/1361-6633/ab28d6}{{\em Rept. Prog. Phys.}
  {\bf 82} (2019) no.~11, 116201}, \href{http://arxiv.org/abs/1806.07396}{{\tt
  arXiv:1806.07396 [hep-ph]}}.

\bibitem{Babu:1997st}
K.~Babu, C.~F. Kolda, and J.~March-Russell, ``{Implications of generalized Z -
  Z-prime mixing},'' \href{http://dx.doi.org/10.1103/PhysRevD.57.6788}{{\em
  Phys. Rev. D} {\bf 57} (1998)  6788--6792},
  \href{http://arxiv.org/abs/hep-ph/9710441}{{\tt arXiv:hep-ph/9710441}}.

\bibitem{Davoudiasl:2012ag}
H.~Davoudiasl, H.-S. Lee, and W.~J. Marciano, ``{'Dark' Z implications for
  Parity Violation, Rare Meson Decays, and Higgs Physics},''
  \href{http://dx.doi.org/10.1103/PhysRevD.85.115019}{{\em Phys. Rev. D} {\bf
  85} (2012)  115019}, \href{http://arxiv.org/abs/1203.2947}{{\tt
  arXiv:1203.2947 [hep-ph]}}.

\bibitem{Dror:2018wfl}
J.~A. Dror, R.~Lasenby, and M.~Pospelov, ``{Light vectors coupled to bosonic
  currents},'' \href{http://dx.doi.org/10.1103/PhysRevD.99.055016}{{\em Phys.
  Rev. D} {\bf 99} (2019) no.~5, 055016},
  \href{http://arxiv.org/abs/1811.00595}{{\tt arXiv:1811.00595 [hep-ph]}}.

\bibitem{Grygier:2017tzo}
{\bf Belle} Collaboration, J.~Grygier {\em et al.}, ``{Search for
  $\boldsymbol{B\to h\nu\bar{\nu}}$ decays with semileptonic tagging at
  Belle},'' \href{http://dx.doi.org/10.1103/PhysRevD.96.091101}{{\em Phys. Rev.
  D} {\bf 96} (2017) no.~9, 091101},
  \href{http://arxiv.org/abs/1702.03224}{{\tt arXiv:1702.03224 [hep-ex]}}.
  [Addendum: Phys.Rev.D 97, 099902 (2018)].

\bibitem{Izaguirre:2016dfi}
E.~Izaguirre, T.~Lin, and B.~Shuve, ``{Searching for Axionlike Particles in
  Flavor-Changing Neutral Current Processes},''
  \href{http://dx.doi.org/10.1103/PhysRevLett.118.111802}{{\em Phys. Rev.
  Lett.} {\bf 118} (2017) no.~11, 111802},
  \href{http://arxiv.org/abs/1611.09355}{{\tt arXiv:1611.09355 [hep-ph]}}.

\bibitem{Aloni:2018vki}
D.~Aloni, Y.~Soreq, and M.~Williams, ``{Coupling QCD-Scale Axionlike Particles
  to Gluons},'' \href{http://dx.doi.org/10.1103/PhysRevLett.123.031803}{{\em
  Phys. Rev. Lett.} {\bf 123} (2019) no.~3, 031803},
  \href{http://arxiv.org/abs/1811.03474}{{\tt arXiv:1811.03474 [hep-ph]}}.

\bibitem{Gavela:2019wzg}
M.~Gavela, R.~Houtz, P.~Quilez, R.~Del~Rey, and O.~Sumensari, ``{Flavor
  constraints on electroweak ALP couplings},''
  \href{http://dx.doi.org/10.1140/epjc/s10052-019-6889-y}{{\em Eur. Phys. J. C}
  {\bf 79} (2019) no.~5, 369}, \href{http://arxiv.org/abs/1901.02031}{{\tt
  arXiv:1901.02031 [hep-ph]}}.

\bibitem{Dolan:2017osp}
M.~J. Dolan, T.~Ferber, C.~Hearty, F.~Kahlhoefer, and K.~Schmidt-Hoberg,
  ``{Revised constraints and Belle II sensitivity for visible and invisible
  axion-like particles},''
  \href{http://dx.doi.org/10.1007/JHEP12(2017)094}{{\em JHEP} {\bf 12} (2017)
  094}, \href{http://arxiv.org/abs/1709.00009}{{\tt arXiv:1709.00009
  [hep-ph]}}.

\bibitem{Bauer:2017ris}
M.~Bauer, M.~Neubert, and A.~Thamm, ``{Collider Probes of Axion-Like
  Particles},'' \href{http://dx.doi.org/10.1007/JHEP12(2017)044}{{\em JHEP}
  {\bf 12} (2017)  044}, \href{http://arxiv.org/abs/1708.00443}{{\tt
  arXiv:1708.00443 [hep-ph]}}.

\bibitem{soton403913}
T.~Kluyver {\em et al.}, ``Jupyter notebooks - a publishing format for
  reproducible computational workflows,'' in {\em Positioning and Power in
  Academic Publishing: Players, Agents and Agendas}, pp.~87--90.
\newblock IOS Press, 2016.
\newblock \url{https://eprints.soton.ac.uk/403913/}.

\bibitem{Hunter:2007}
J.~D. Hunter, ``Matplotlib: A 2d graphics environment,''
  \href{http://dx.doi.org/10.1109/MCSE.2007.55}{{\em Computing in Science \&
  Engineering} {\bf 9} (2007) no.~3, 90--95}.

\bibitem{numpy}
T.~Oliphant, ``{NumPy}: A guide to {NumPy}.'' Usa: Trelgol publishing, 2006--.
\newblock \url{http://www.numpy.org/}.

\bibitem{lukas_2018_1217032}
L.~Heinrich, ``lukasheinrich/pylhe v0.0.4,'' Apr., 2018.
\newblock \url{https://doi.org/10.5281/zenodo.1217032}.

\bibitem{Rodrigues:2019nct}
E.~Rodrigues, ``{The Scikit-HEP Project},'' in {\em {23rd International
  Conference on Computing in High Energy and Nuclear Physics (CHEP 2018) Sofia,
  Bulgaria, July 9-13, 2018}}.
\newblock 2019.
\newblock
\href{http://arxiv.org/abs/1905.00002}{{\tt arXiv:1905.00002
  [physics.comp-ph]}}.
\newblock

\bibitem{Ball:2004ye}
P.~Ball and R.~Zwicky, ``{New results on $B \to \pi, K, \eta$ decay formfactors
  from light-cone sum rules},''
  \href{http://dx.doi.org/10.1103/PhysRevD.71.014015}{{\em Phys. Rev. D} {\bf
  71} (2005)  014015}, \href{http://arxiv.org/abs/hep-ph/0406232}{{\tt
  arXiv:hep-ph/0406232}}.

\bibitem{Ball:2004rg}
P.~Ball and R.~Zwicky, ``{$B_{d,s} \to \rho, \omega, K^*, \phi$ decay
  form-factors from light-cone sum rules revisited},''
  \href{http://dx.doi.org/10.1103/PhysRevD.71.014029}{{\em Phys. Rev. D} {\bf
  71} (2005)  014029}, \href{http://arxiv.org/abs/hep-ph/0412079}{{\tt
  arXiv:hep-ph/0412079}}.

\end{thebibliography}\endgroup

\end{document}